\documentclass[10pt]{article}
\usepackage{latexsym,amsmath,amssymb,amsthm,amsfonts,graphicx}
\numberwithin{equation}{section}
\numberwithin{figure}{section}
\author{Roy H. Goodman%
\thanks{Department of Mathematical Sciences, New Jersey Institute of
  Technology, Newark, NJ 07102}\hspace{.2cm}
Richard Haberman%
\thanks{Department of Mathematics, Southern Methodist University, Dallas, 
TX 75275}
}
\newcommand{\nc}{\newcommand} 
\nc{\diff}[2]{\frac{d #1}{d #2}}
\nc{\abs}[1] {\lvert #1 \rvert}

\nc{\e}{\epsilon} 
\DeclareMathOperator{\sech}{sech}
\DeclareMathOperator{\intg}{int}
\DeclareMathOperator{\fr}{fr}
\nc{\intinf}{\int_{-\infty}^{\infty}}
\def\beq#1\eeq{\begin{equation}#1\end{equation}}
\nc{\vin}{v_{\rm in}}
\nc{\vout}{v_{\rm out}}
\nc{\vc}{v_c}
\nc{\half}{\frac{1}{2}}

\title{Interaction of sine-Gordon kinks with defects: The two-bounce resonance}
\begin{document}

\maketitle 

\begin{abstract}
{A model of soliton-defect interactions in the sine-Gordon equations is
studied using singular perturbation theory.  Melnikov theory is used to
derive a critical velocity for strong interactions, which is shown to be
exponentially small for weak defects.  Matched asymptotic expansions for
nearly heteroclinic orbits are constructed for the initial value problem,
which are then used to derive analytical formulas for the locations of the
well known two- and three-bounce resonance windows, as well as several other
phenomena seen in numerical simulations.}
\end{abstract}

\section{The two-bounce resonance}

\begin{figure}
\begin{center}
\includegraphics[width=3in]{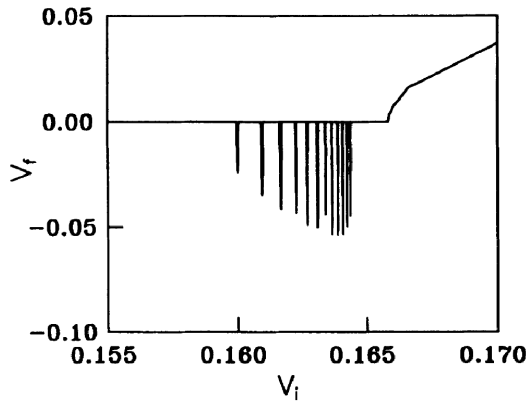}
\caption{The output vs. input velocities of sine-Gordon solitons interacting
  with a delta-well defect, from~\cite{FKV:92}, reprinted with permission.}
\label{fig:fkv}
\end{center}
\end{figure}

The two-bounce resonance is a phenomenon displayed by many non-integrable
systems in which a solitary wave interacts either with another solitary wave
or else with a localized defect in the medium through which it propagates.
Fei, Kivshar, and V\'azquez study the two-bounce resonance in the sine-Gordon
equation perturbed by a localized nonlinear defect~\cite{FKV:92}.  
\begin{equation}
u_{tt}-u_{xx}+\sin{u} = \e \delta(x) \sin{u}. \label{eq:sg}
\end{equation}
Kink solitons are initialized propagating (numerically) toward a defect 
with velocity $v_{\rm i}$ and allowed to interact
with the defect.   Then one of two things might happen: either the
soliton is trapped and comes to rest at the defect location, or else it 
escapes and propagates away at finite speed $v_{\rm f}$.  
(The soliton cannot be destroyed by the interaction because it is defined
by its boundary conditions at infinity.)  They find that there exists a critical velocity $v_{\rm c}$. 
Kink solitons with initial velocity greater than $v_{\rm c}$ pass by the defect.   Most solitons with initial speeds below
the $v_{\rm c}$ are trapped, remaining at the defect for all times after
the interaction time.  However, there exist bands of initial velocities, known
as resonance windows, for which the kink is reflected by the defect, rather
than being trapped.  This is summarized in Figure~\ref{fig:fkv}, taken from
their paper.

A phenomenological explanation for this phenomenon (in the context of
kink-antikink interactions in nonlinear Klein-Gordon equations) was given by
Campbell et.\ al.~\cite{CSW:83,PC:83,CP:86, CPS:86} in a series of of papers.
They use very elegant physical reasoning to argue that the resonance
windows are due to a resonant interaction between the movement of the
kink-antikink pair in an effective potential, and shape modes oscillating
about the kinks.  Fei et.\ al.\ give an analysis of the two-bounce resonance
phenomenon which relies on a variational approximation, which reduces the
sine-Gordon PDE to a pair of second order ODE, and use a similar argument to
find the resonance windows.  Both these studies make the assumptions that the
resonance takes a certain form, dependent on unknown constants, and use a mix
of physical reasoning and statistical data fitting to find these constants.

An inspiration for the present work comes from one of the authors' previous studies
of the trapping of gap solitons in Bragg grating optical fibers with defects~\cite{GSW:02}.  In that 
study, sufficiently slow solitons in certain parameter regimes were captured by localized defects.  This previous work does not offer a mechanism to explain the existence of a critical velocity for soliton
capture, which we are now able to explain for the simpler model problem discussed here.  The two bounce resonance phenomenon is also seen by Tan and Yang in simulations of vector solitons collisions in birefringent optical fibers~\cite{TY:01}.

The aim of the current paper is to make mathematically precise the physical
reasoning of the previous studies of the two-bounce resonance, in a way that
does not rely on statistical inference.  We analyze the variational ODE model derived in~\cite{FKV:92}. We use the methods of singular perturbation theory to match a nonlinear 
saddle to nearly heteroclinic orbits in a manner similar to that previously used by Haberman~\cite{H:01, H:02} and Diminnie and Haberman~\cite{DH:02, DH:03}.  The
critical velocity is determined via a Melnikov integral and the location of
the resonance windows arises naturally due to a matching condition in the
expansion.  Intriguingly, finding the critical velocity requires that we make
use of terms which are small beyond all orders in $\e$ in the matched asymptotic
expansion, as was done, notably, by Kruskal and Segur~\cite{KS:91}, and by
many others.

Other work on soliton dynamics in perturbed sine-Gordon equations is summarized by Scott~\cite{Scott}.  In this approach, an ordinary differential equation is derived for the evolution of the Hamiltonian, which can then be related to the soliton's velocity.  McLaughlin and Scott~\cite{MS:78} study
a damped and driven sine-Gordon system modeling a Josephson junction and find a unique limiting velocity for solitons under that perturbation.  The fundamental difference between their system and ours is the presence of the localized defect mode, which must be included in the reduced system.

The paper is laid out as follows.  In section~\ref{sec:model} we introduce a
system of ordinary differential equations that models equation~\eqref{eq:sg},
and show the results of numerical simulations of the model.  In
section~\ref{sec:velocity}, we determine the critical velocity separating
captured kinks from those that pass by the defect.  In
section~\ref{sec:near_inf}, we derive formulas that are valid in a
neighborhood of $\abs{X}=\infty$.  These are used in
section~\ref{sec:construction} where we construct matched asymptotic expansions to
solutions satisfying the 2-bounce resonance.  We find the sequence of
velocities defining the resonance windows, as well as formulas for the window
widths.  We also find locations of 3-bounce resonance windows and approximations
for the general initial value problem. In
section~\ref{sec:numerics}, we demonstrate the validity of this approach by
comparing the formulas derived in the previous two sections with the results
of numerical simulations.  We summarize and include a more general discussion
in section~\ref{sec:summary}.

\section{The Variational Approximation}
\label{sec:model} 

Following Fei et al.~\cite{FKV:92}, we consider a sine-Gordon model with a
localized impurity at the origin, given by equation~\eqref{eq:sg}.
In the absence of any impurity, i.e. $\e=0$, the sine-Gordon equation has the
well-known family of kink solutions parameterized by velocity $v$:
\begin{equation}
u_k(x,t)= 4 \tan^{-1} \exp{\left( (x - v t -x_0)/\sqrt{1-v^2}\right)}.
\end{equation}
If we consider the system with an impurity, then solutions of small amplitude
approximately satisfy the linear equation:
\begin{equation}
u_{tt}-u_{xx}+u = \e \delta(x) u,
\label{eq:linear_delta}
\end{equation}
which, for $0 < \e <2$,  has standing wave solutions
$$
u_{\rm im}(x,t) = a(t) e^{-\e\abs{x}/2},
$$
where $a(t) = a_0 \cos(\Omega t+\theta_0)$ and 
\begin{equation}
\Omega = \sqrt{1-\e^2/4}.
\label{eq:Omega}
\end{equation}
Fei, Kivshar, and V\'{a}zquez~\cite{FKV:92} study the interaction of the kink
and defect modes using a variational approximation to derive a set of
equations for the evolution of the kink position $X$, and the defect mode
amplitude $a$.  An excellent review of the use of variational approximations
in nonlinear optics is given by Malomed~\cite{M:02}.  To derive the
approximate equations, they substitute the ansatz
\begin{equation}
u=u_k + u_{\rm im}= 4 \tan^{-1} \exp{ (x - X(t))}+ a(t) e^{-\e\abs{x}/2}
\label{eq:ansatz1}
\end{equation}
into the Lagrangian of~\eqref{eq:sg}
\begin{equation}
\label{eq:Lag}
L =\intinf  \left( \frac{1}{2}u_t^2 -\frac{1}{2}u_x^2 
-[1-\e\delta(x)](1-\cos{u}) \right) dx.
\end{equation}
Here $X$ replaces $x_0 + Vt$, and $a$ and $X$, the parameters characterizing
the approximate solution~\eqref{eq:ansatz1}, are regarded as unknown functions
of $t$. It is assumed that $a$ and $\e$ are small enough that many cross-terms
can be neglected. Thus, in calculating the effective Lagrangian, all terms
produced via overlap of the two modes are neglected, excepting those which
include the defect potential $\delta(x)$.  This is equivalent to assuming that
the dominant means of interaction between the two modes is via the defect.
Evaluating the spatial integrals of~\eqref{eq:Lag}, an effective Lagrangian
$L_{\rm eff}(X, a, \Dot X, \Dot a)$ is obtained~\cite{FKV:92}:
\begin{equation}
\label{eq:effLag} 
L_{\rm eff} =4 {\Dot X}^2 + \frac{1}{\e}(\Dot a^2 - \Omega^2 a^2)
 - \e U(X) - \e aF(X),
\end{equation}
where
\begin{align*}
 U(X) &= -2 \sech^2(X); \\
 F(X) &= -2 \tanh(X)\sech(X).
\end{align*}
The corresponding evolution equations are then given by the classical
Euler-Lagrange equations for~\eqref{eq:effLag}:
\begin{subequations}
\label{eq:model}
\begin{align}
 8 \Ddot X + \e U'(X) + \e aF'(X) &=0; \label{eq:model1}\\
 \Ddot a + \Omega^2 a + \frac{\e^2}{2} F(X) &=0. \label{eq:model2}
\end{align}
\end{subequations}
This system has also been studied in~\cite{GHW:02}.  Note that the system
conserves the Hamiltonian 
\beq 
H = 4 {\Dot X}^2 + \frac{1}{\e}(\Dot a^2 + \Omega^2 a^2) 
+ \e U(X) + \e aF(X) \label{eq:H} 
\eeq 
and that as $\abs{X} \to \infty$, $U\to 0$ and $F\to 0$ exponentially.  The energy is thus
asymptotically positive definite, and must be partitioned between $X$ and $a$
when the soliton is far from the defect.

This system corresponds to a particle $X$ moving in an attractive potential
well $\e U(X)$ exponentially localized in a neighborhood of zero, coupled to a
harmonic oscillator $a$ by an exponentially localized term $\e aF(X)$.  Note
that this model inherits many properties from the sine-Gordon system.  $U(X)$
and $F(X)$ decay for large $\abs{X}$, so that when $\abs{X}$ is large
$\Ddot X \approx 0$ and the kink may propagate at any constant speed,
independent of the impurity mode $a$, which oscillates at its characteristic
frequency $\Omega$.  When $X$ becomes small, the two equations become coupled
and the kink may exchange energy with the impurity mode.

The variational method, while popular in the study of nonlinear optics, may
contain significant pitfalls.  First, it depends on the investigator finding
an appropriate ansatz, as is done in equation~\eqref{eq:ansatz1}.  Second,
even if the ansatz is chosen to be an exact representation of the initial
data, there is no guarantee given by the method that the solution at a later
time is well represented by an approximation of this form.  Thus, one must
carefully show that solutions of the full PDE system are well approximated by
the ansatz.

\begin{figure}
\begin{center}
\includegraphics[width=4in]{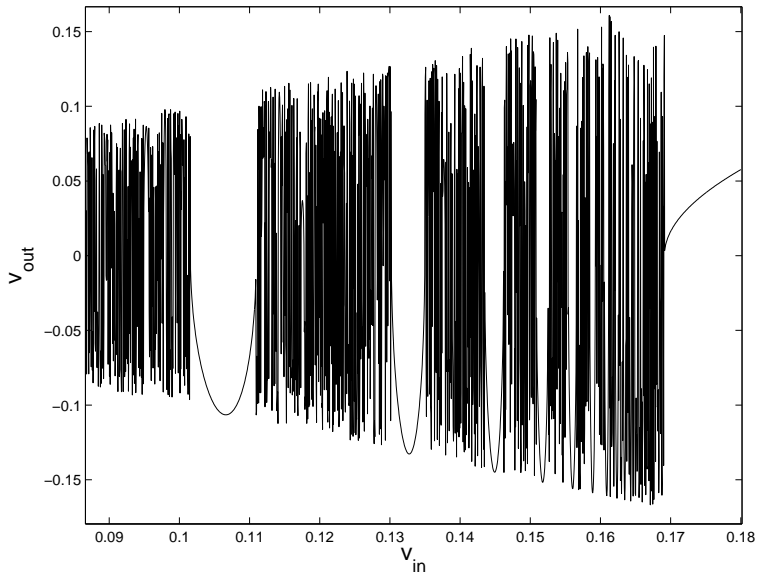}
\caption{The analog of figure~\ref{fig:fkv} for the ODE~\eqref{eq:model}, with
  $\e=0.5$.}
\label{fig:v_in_out}
\end{center}
\end{figure}

Figure~\ref{fig:fkv} should be compared to figure~\ref{fig:v_in_out}.  The
former plots the output versus input velocities for the full PDE, as computed
in~\cite{FKV:92}.  It shows a critical velocity $v_{\rm c}\approx 0.166$, and a
finite number of resonance windows of decreasing width as $v \nearrow v_{\rm c}$.
In between these resonance windows, incoming solutions are trapped.  For
speeds slightly above $v_{\rm c}$, it appears that $v_{\rm f}=
O((v_{\rm i}-v_{\rm c})^{\frac{1}{2}})$.  The latter shows the same experiment for the
ODE.  This shows a critical velocity $v_{\rm c}\approx 0.17$, in reasonable
agreement with the PDE dynamics, a sequence of reflection windows, 
and a square-root profile just to the right of $v_{\rm c}$.  There are a few major
differences between the two numerical experiments.
The first is that the PDE dynamics show only a finite number of resonance
windows, while the number of resonance windows for the ODE dynamics will be
shown below to be infinite.  Second, the regions between the resonance windows
do not usually give rise to trapped solutions.  It was shown in~\cite{GHW:02}
that almost all solutions have nonzero
$\vout$.  This is because the variational ODEs are Hamiltonian, and a variant
of the Poincar\'e recurrence theorem implies that the probability that a
solution is trapped is zero.  Also note, that the exit speed in the resonance windows for the PDE computation is significantly smaller than the input speed, while for the ODE, the $\vout=-\vin$ at the center of the resonance windows.
The variational ansatz~\eqref{eq:ansatz1}
ignores energy that is lost via transfer to radiation modes.
In~\cite{GHW:02}, a dissipative correction to~\eqref{eq:model} is derived that
takes this into account. This eliminates most of the sensitive dependence of
$\vout$ on $\vin$ and replaces the chaotic regions with trapping regions.
Nonetheless, we believe the Hamiltonian ODE~\eqref{eq:model} displays the fundamental features, if not the exact details, of the two-bounce resonance.

\begin{figure}
\begin{center}
\includegraphics[width=2in]{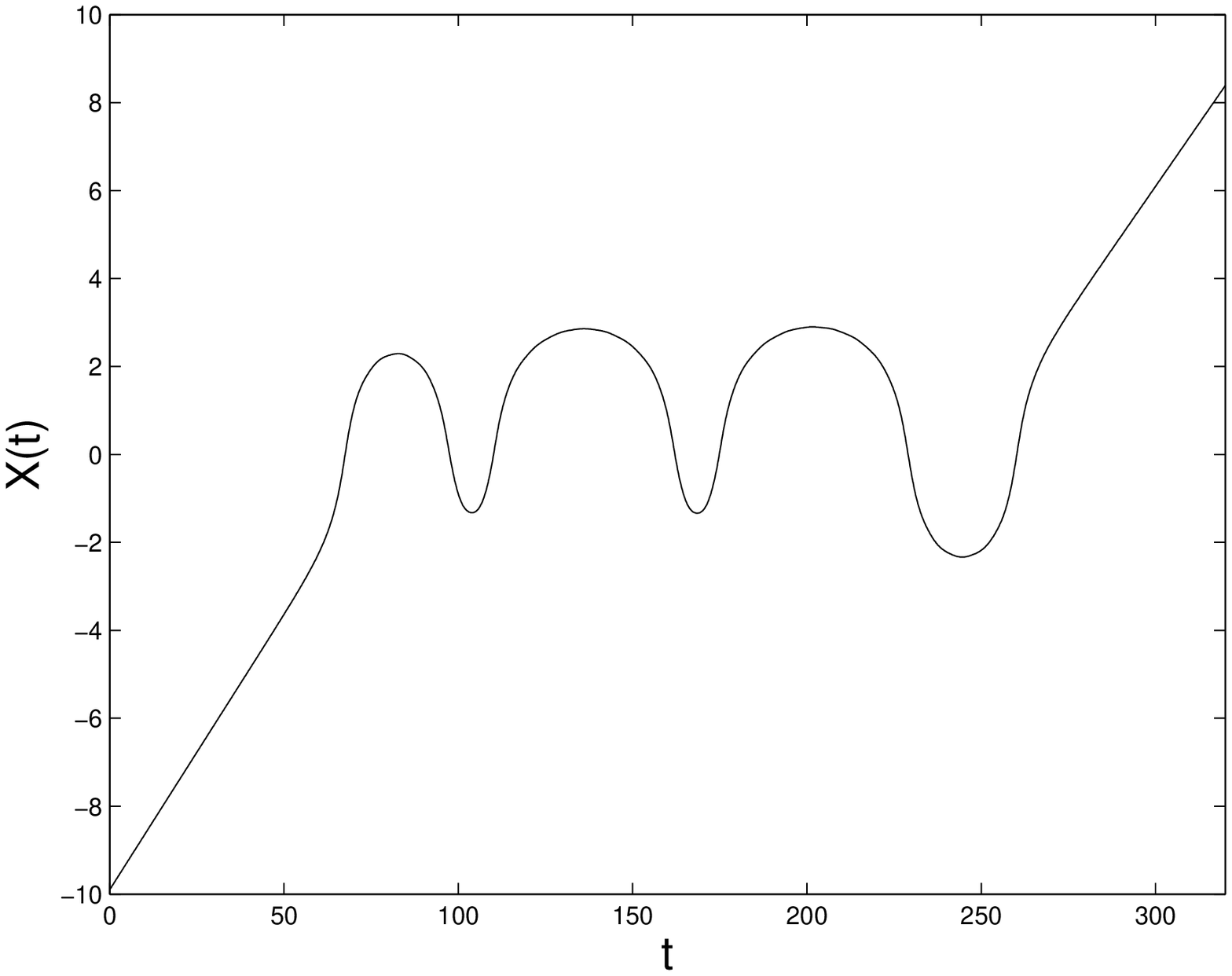}
\includegraphics[width=2in]{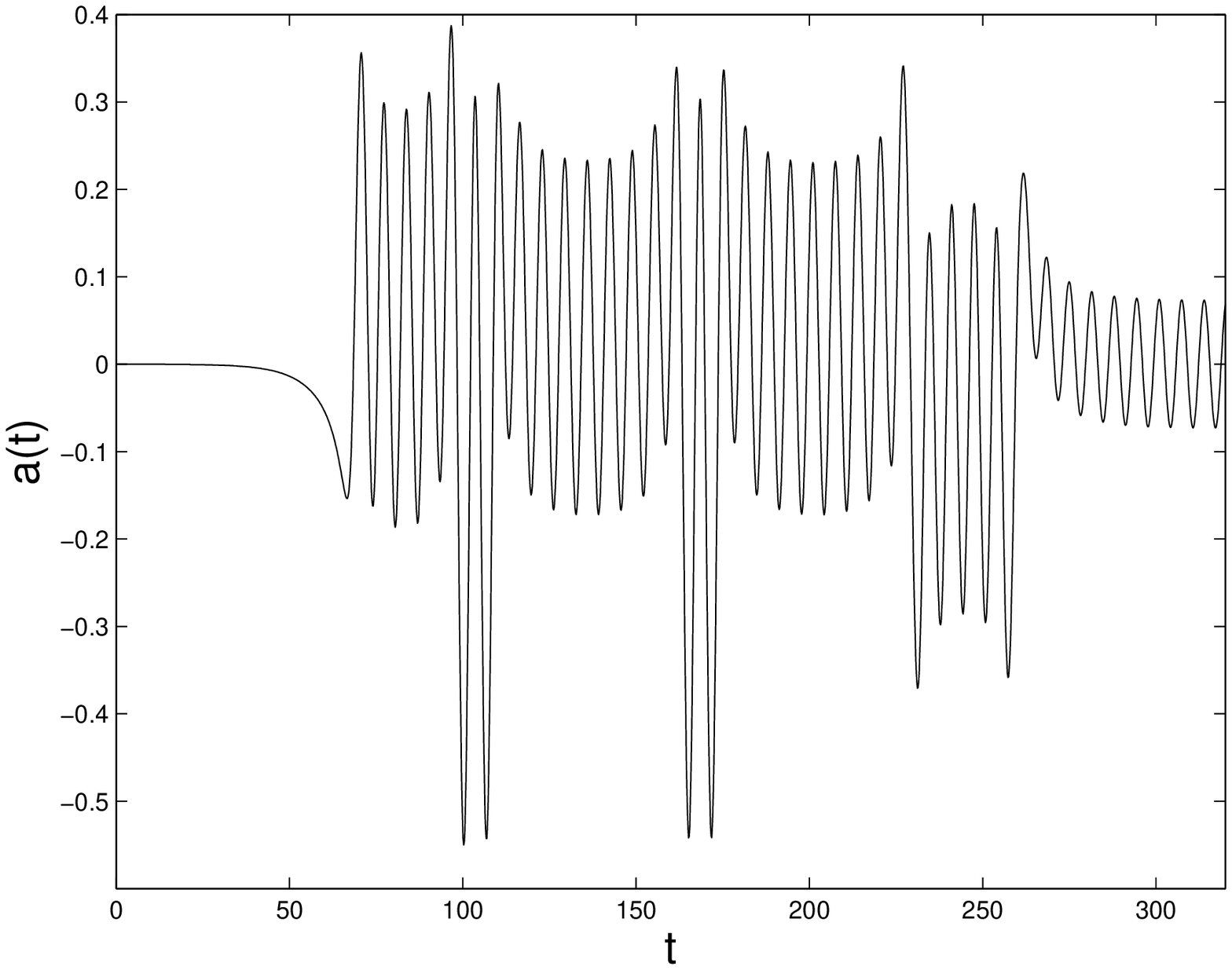}
\caption{X(t) and a(t) for the numerical experiment with $\e=0.5$ and
  $\vin=0.125$.}
\label{fig:general_solution}
\end{center}
\end{figure}

\begin{figure}
\begin{center}
\includegraphics[width=2in]{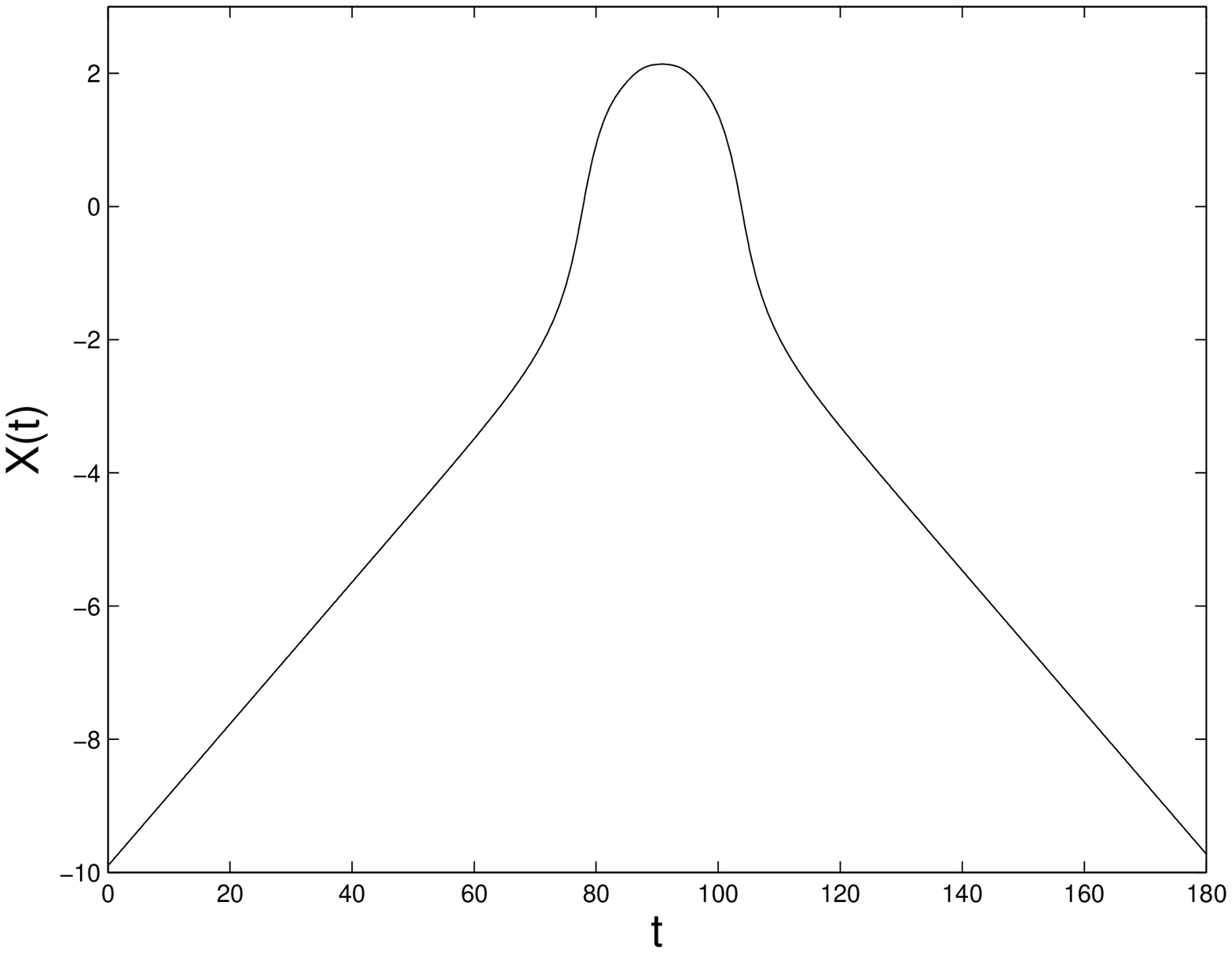}
\includegraphics[width=2in]{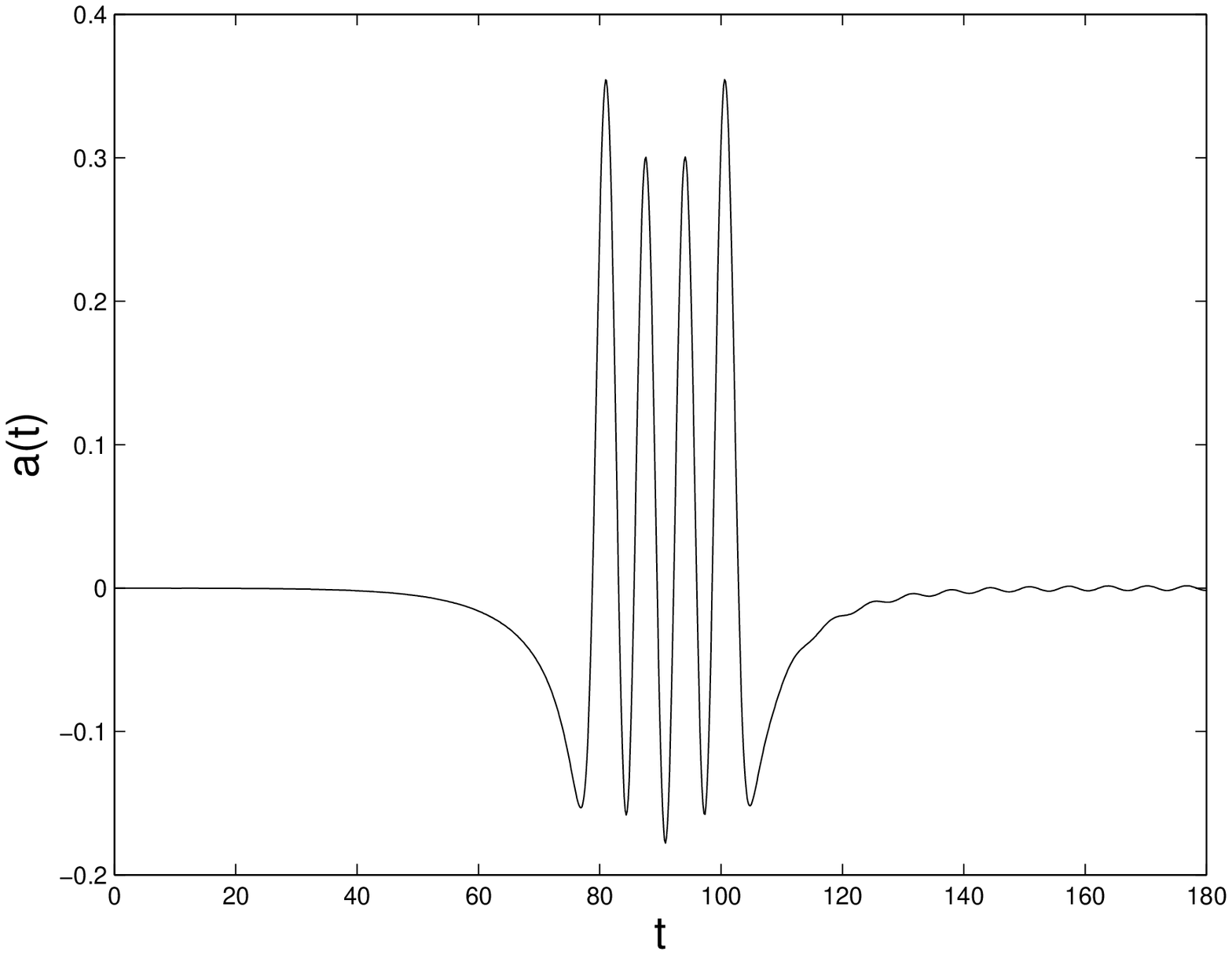}
\caption{X(t) and a(t) for the numerical experiment with $\e=0.5$ and
  $\vin=0.10645$, showing the 2-4 resonance.}
\label{fig:2-4bounce}
\end{center}
\end{figure}

\begin{figure}
\begin{center}
\includegraphics[width=2in]{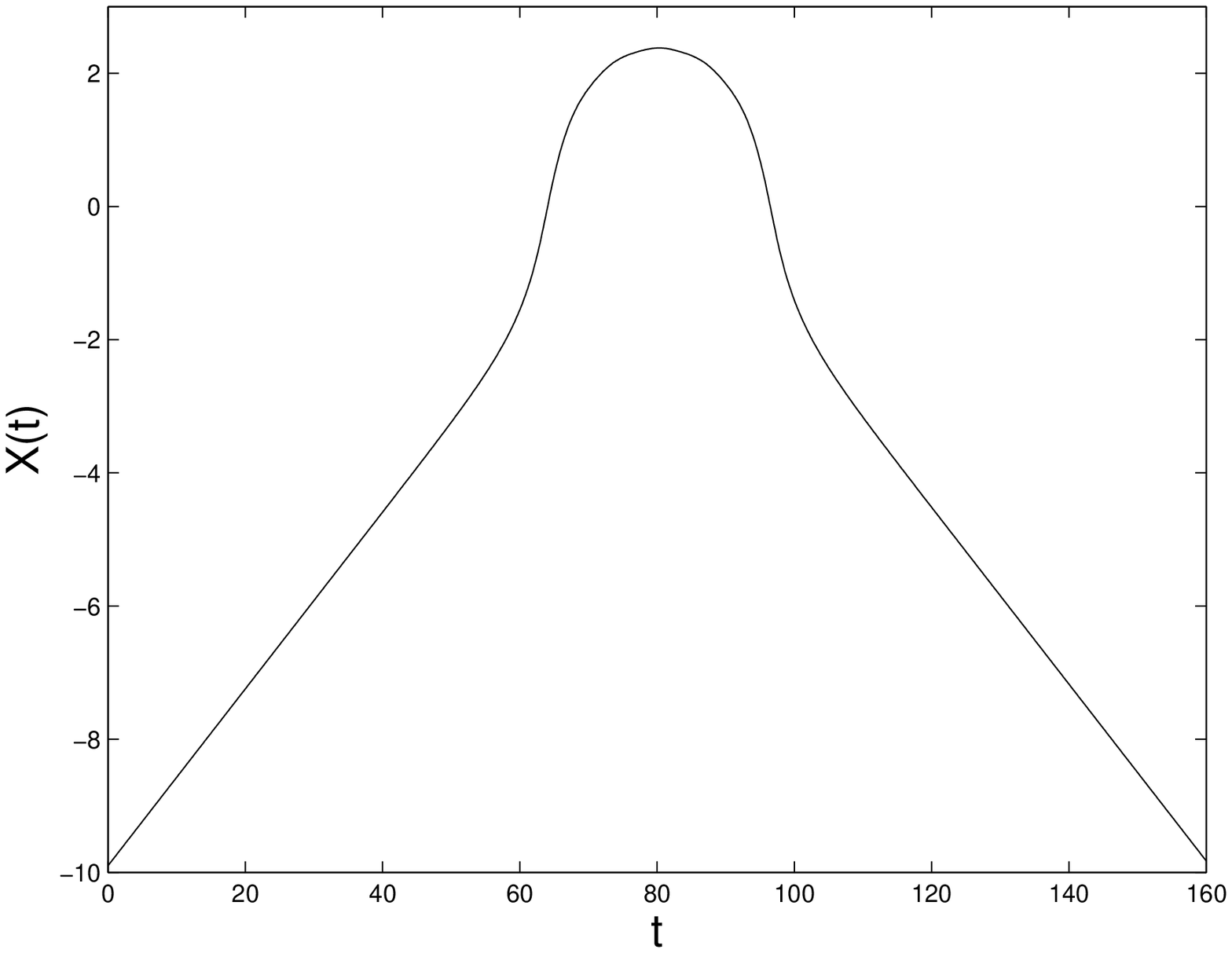}
\includegraphics[width=2in]{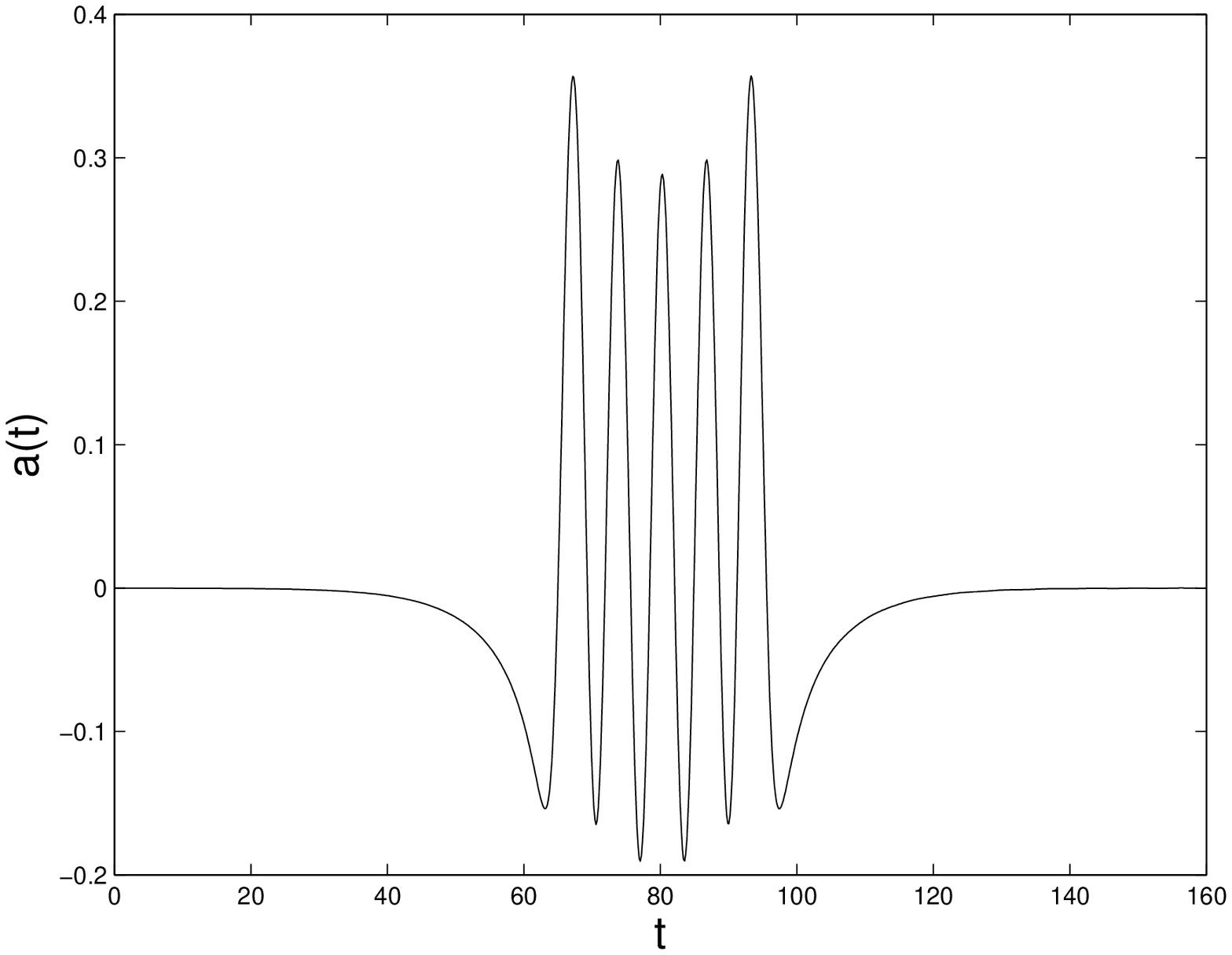}
\caption{X(t) and a(t) for the numerical experiment with $\e=0.5$ and
  $\vin=0.1327$, showing the 2-5 resonance.}
\label{fig:2-5bounce}
\end{center}
\end{figure}

We now describe the structure of individual solutions to the
ODE~\eqref{eq:model}.  The numerical experiments were performed with initial
conditions
$$ X(0) = -12; \; \Dot X(0) = \vin>0; \; a(0)=0; \; \dot a(0)=0. $$ 
For a general value of $\vin<\vc$, $X(t)$
comes in at constant speed, speeds up near zero, slows down as it approaches $+\infty$, oscillates
back and forth a few times, then emerges and heads off in either direction
with finite velocity $\vout$, with $\abs{\vout} \le \vin$. The harmonic
oscillator $a(t)$, at first grows monotonically, and then begins oscillating,
interrupted by a sequence of jumps in its amplitude and phase, before settling
down to a steady oscillation as $X\to\infty$; see
figure~\ref{fig:general_solution}.  This includes the $\vin$ in the two-bounce resonance
windows, in which the behavior is 
simpler: $X(t)$ approaches plus infinity, turns around, and heads back off to minus
infinity and $a(t)$ grows, oscillates a finite number of time, and then
shrinks again.  At the very bottom of the resonance window (actually at a
point tangent to the line $\vout=-\vin$ in figure~\ref{fig:v_in_out}),
$a(t)$ actually returns all its energy to $X(t)$, so that $\lim_{t\to
  \infty}a(t) = 0$ and $\vout = - \vin$.  In each successive window, the
$a(t)$ undergoes one more oscillation than in the window to its left, with
$n_{\rm min}(\e)$ oscillations in the leftmost window.  This number increases
quickly as $\e\searrow 0$.  For example, when $\e=0.5$, $a(t)$ undergoes 4
oscillations for $\vin$ in the leftmost window, 5 in the next window, etc.; see
figures~\ref{fig:2-4bounce} and~\ref{fig:2-5bounce}.  The phrase ``2-bounce
resonance'' was coined in~\cite{CSW:83} and refers to the fact that the kink
comes in contact with the defect twice; e.g.\ in figure~\ref{fig:2-4bounce},
these would be at about $t=80$ and $t=100$ when $X=0$.  It is during the
``bounces'' that the kink is in contact with the defect and exchanges energy with
the defect mode.  During the first interaction the soliton gives up energy to
the defect mode and is trapped, and in the second, the energy is returned, and
the soliton resumes propagating.  We generalize this name to the 2-$n$ bounce
resonance, where $n$ denotes the number of complete oscillations of $a(t)$.
It is possible to find in the simulations higher resonances, where the soliton
interacts with the defect three or more times, before its energy is returned
and it resumes propagating.  These resonance windows are naturally much
narrower.  Interspersed between the reflection and transmission windows is a set of initial conditions of measure zero in which the solutions are chaotic and $X(t)$ remains bounded 
for all time

\begin{figure}
\begin{center}
\includegraphics[width=3in]{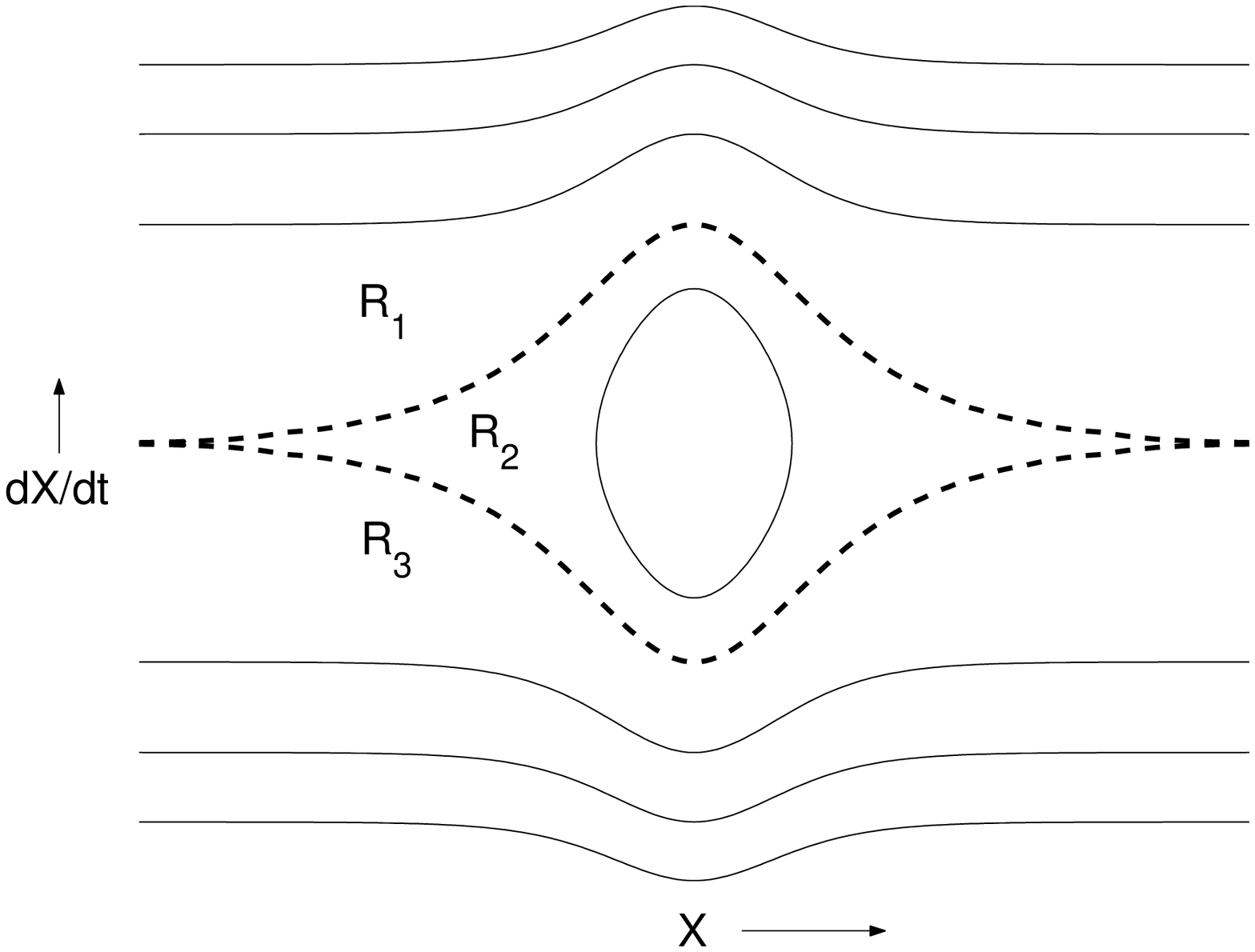}
\caption{The phase plane of the uncoupled $X$ dynamics, divided into three
  regions by a pair of degenerate heteroclinic orbits.}
\label{fig:phaseplane}
\end{center}
\end{figure}

\begin{figure}
\begin{center}
\includegraphics[width=2in]{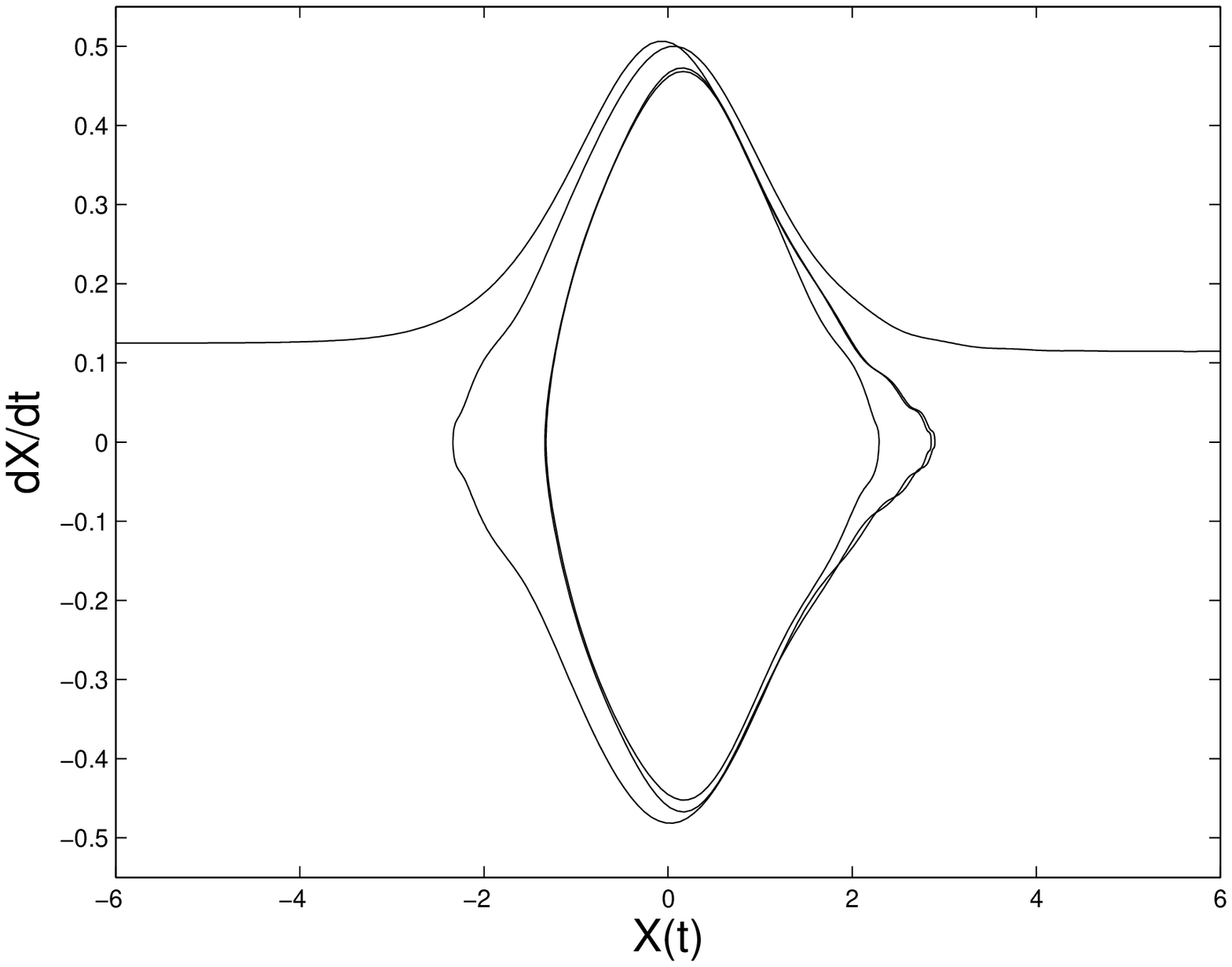}
\includegraphics[width=2in]{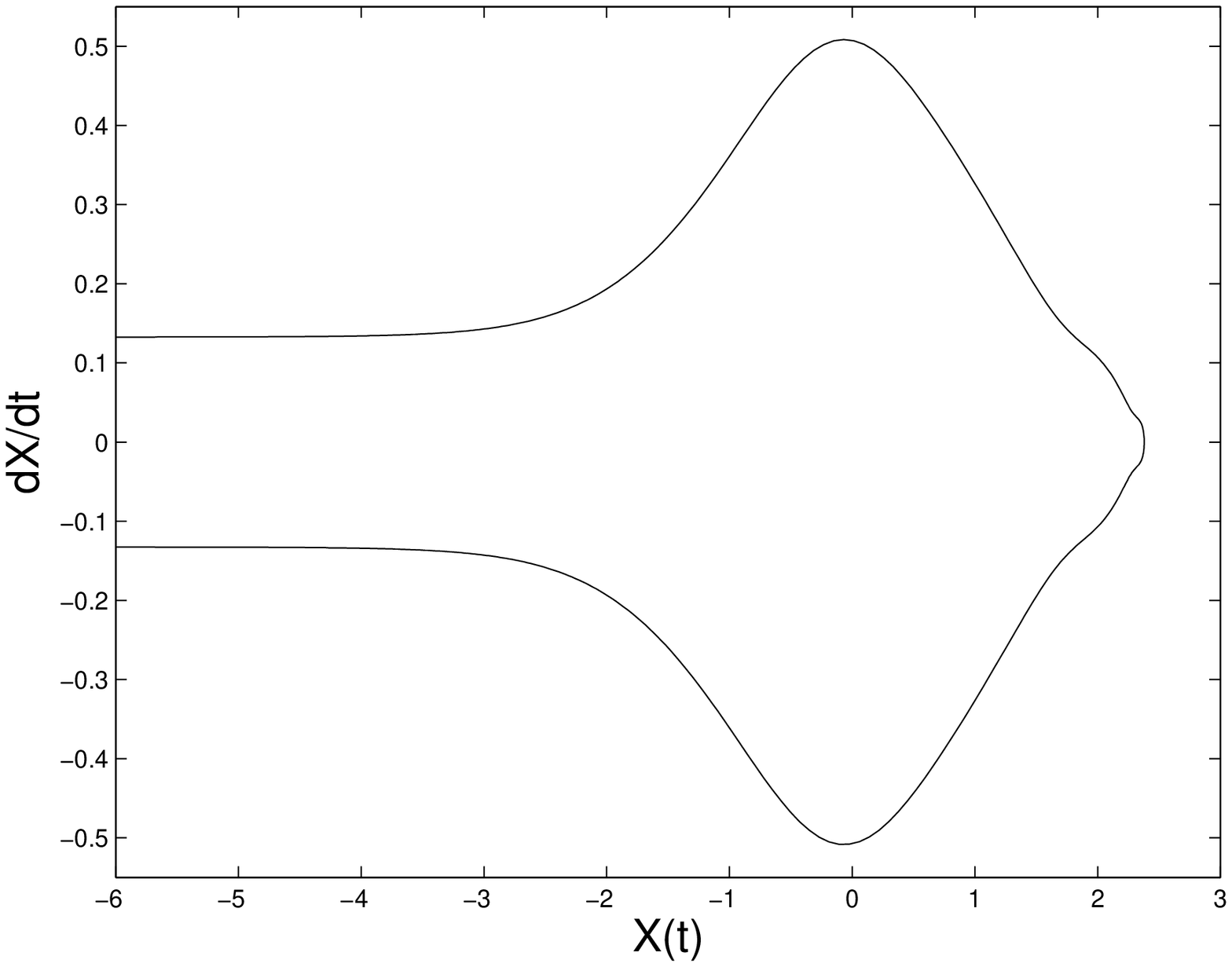}
\caption{Projections into $(X,\dot X)$ plane of the solutions shown in
  figure~\ref{fig:general_solution} (left) and figure~\ref{fig:2-5bounce}
  (right).}
\label{fig:phase_projections}
\end{center}
\end{figure}

It is helpful to look at projections of the solutions in the $(X,\dot X)$
phase space.  If we ignore the term $\e a F'(X)$ in~\eqref{eq:model1}, the
simplified system has an elliptic fixed point at $(0,0)$ and degenerate
saddle-like fixed points at $(\pm \infty, 0)$, connected by a pair of
heteroclinic orbits, which split the phase space into three regions, as is
shown in figure~\ref{fig:phaseplane}.  In region $R_{1}$ (respectively
$R_{3}$), solutions move right (respectively left) along trajectories that
asymptote to horizontal lines for large $\abs{X}$.  Solutions in region
$R_{2}$ oscillate clockwise, remaining bounded for all time.  When the
coupling to $a(t)$ is restored, these trajectories are no longer invariant,
and the solution may cross over the separatrices.  A typical solution starting
in region $R_{1}$ will cross over the separatrix, oscillate inside $R_{2}$
several times, then exit to either region $R_{1}$ or $R_{3}$; as is shown in
the first graph of figure~\ref{fig:phase_projections}.  In a 2-bounce
solution, $X(t)$ must first cross from $R_{1}$ to $R_{2}$, undergo half an
oscillation, and then cross into $R_{3}$ and propagate back toward $-\infty$;
as is seen in the second graph of figure~\ref{fig:phase_projections} for an
illustration.

\section{Determination of the critical velocity}
\label{sec:velocity}
To compute the critical velocity~$\vc$, we will use a Melnikov
computation~\cite{GH:83, M:63}.  Essentially, we write down the time rate of
change of the energy contained in $X(t)$, and integrate this over a separatrix
orbit to find the total energy transferred away from $X$ as it travels from
$-\infty$ to $+\infty$.  Then, if the initial energy is greater than the
energy loss, then $X$ reaches $+\infty$.  If the energy is less, than the
trajectory crosses the separatrix and turns around.

We rescale the time variable $t \to \sqrt{\frac{\e}{2}} t$.  Under this
scaling, the equations become:
\begin{subequations}
\label{eq:scaledmodel}
\begin{align}
 4 \Ddot X +  U'(X) + aF'(X) &=0; \label{eq:model_a}\\
 \Ddot a + \lambda^2 a + \e F(X) &=0 \label{eq:model_b}
\end{align}
\end{subequations}
where 
$$\lambda^2 = \frac{2}{\e}-\frac{\e}{2}.$$
This removes the explicit $\e$-depence from~\eqref{eq:model_a} and fixes
the leading-order time scale.

We consider the initial value problem defined by~\eqref{eq:scaledmodel}
together with the ``initial condition'' that as $t\to-\infty$,
\begin{equation}
X \to -\infty; \; \dot X \to V \; a \to 0 \; \dot a \to 0.
\label{eq:init_cond}
\end{equation}
Because~\eqref{eq:scaledmodel} is autonomous, this is insufficient to specify 
a unique solution, and we should append the condition that as $t\to -\infty$,
$$
X \sim X_{0}-V t.
$$
When $\e=0$, $a = \dot a =0$ defines an invariant subspace ${\cal P}_{0 }$
of~\eqref{eq:scaledmodel} with trajectories confined to lie on surfaces along
which the energy
\begin{equation}
 E= 2 \Dot X^2 + U(X)
\label{eq:E_energy}
\end{equation} 
is constant.

As seen in figure~\ref{fig:phaseplane}, the unperturbed $X$-phase space
features bounded periodic orbits for $E<0$, unbounded orbits which tend to a
finite velocity at $\abs{t}\to\infty$ for $E>0$ and separatrix orbits with
$E=0$ along which $\Dot X\to 0$ as $\abs{X} \to \infty$.  Along this heteroclinic orbit
\beq
X= \pm\sinh^{-1}{(t-t_{1})} \label{eq:separatrix}
\eeq
where $t_{1}$ is the ``symmetry time'' of the orbit.  In the calculation that follows, we 
will set $t_{1}=0$ for ease of notation.  We will need to include nonzero $t_{1}$ later, 
and will reintroduce it at key locations in the computation.

When $\e>0$, ${\cal P}_{0}$ ceases to be invariant, and energy is transferred
from $X$ to $a$.  Because the coupling term $F(X)$ decays exponentially,
almost all the energy exchange takes place when $X$ is small.  This justifies calculating the
change of energy along the separatrix, because very little of the change of energy is due to the
tails.  We now compute
the change in energy for small values of $\e$, as $X$ travels from $-\infty$
to $+\infty$.  We do this using a Melnikov integral.

Using equations~\eqref{eq:E_energy} and~\eqref{eq:model_a}, the time derivative
of the energy in $E$ is
\begin{equation}
\begin{split}
\diff{E}{t} & = ( 4 \Ddot X +  U'(X))\Dot X \nonumber \\
 &= - aF'(X)\Dot X .
\end{split}
\end{equation}

Integrating this over the separatrix orbit yields the approximate total loss of energy
of the soliton over the trajectory in the form of a Melnikov integral:
\begin{equation*}
\begin{split}
\Delta E &= \intinf \diff{E}{t} dt \\
&= -\intinf  F'(X(t))\Dot X(t) a(t) dt .
\end{split}
\end{equation*}

Plugging the various formulae into the separatrix~\eqref{eq:separatrix} (using
the plus signs for right-moving trajectories and allowing $t_{1}=0$, which
does not effect this calculation):
\begin{gather*}
F= - 2 \sech X \tanh X = \frac{-2t}{1+t^2}; \\
F' = -4 \sech^3 X + 2 \sech X 
   = \frac{-4}{(1+t^2)^{3/2}} +\frac{2}{(1+t^2)^{1/2}} ;\\
\Dot X = \sech X = (1+t^2)^{-1/2}.
\end{gather*}
This gives the Melnikov integral
\beq
\begin{split}
\Delta E 
=-\intinf\Bigl(\frac{-4}{(1+t^2)^{2}} +\frac{2}{1+t^2}\Bigr) a(t)dt. \\
\label{eq:melnikov}
\end{split}
\eeq 

We evaluate $\Delta E$ by first computing $a(t)$ and then using this in
equation~\eqref{eq:melnikov}.
Using initial condition~\eqref{eq:init_cond}, we
may solve for $a$ by variation of parameters:
\begin{equation}
\begin{split}
a &= \frac{\e}{\lambda}\cos{\lambda t} \int_{-\infty}^{t} F(X(\tau))
\sin{\lambda \tau} d\tau - \frac{\e}{\lambda}\sin{\lambda t} 
\int_{-\infty}^{t}F(X(\tau)) \cos{\lambda \tau} d\tau \\
&= -\frac{\e}{\lambda}\int_{-\infty}^{t} F(X(\tau)) 
\sin{\lambda(t-\tau)}d\tau\\
&= \frac{2\e}{\lambda}\int_{-\infty}^t \sin{\lambda(t-\tau)}
\frac{\tau}{1+\tau^2} d\tau.
\label{eq:a_integral}
\end{split}
\end{equation}
In fact, only the even component of $a(t)$ will be needed to evaluate $\Delta E$. 
This is given by
\beq
a_{\rm e}=\frac{\e}{\lambda}\intinf  \sin{\lambda(t-\tau)}
\frac{\tau}{1+\tau^2} d\tau.
\label{eq:a-even}
\eeq
This may be evaluated by introducing the complex exponential and closing the integral in the lower half $\tau$-plane, 
which gives a contribution from the pole at $\tau=-i$:
\beq
a_{\rm e}= -\frac{\e\pi e^{-\lambda}} {\lambda} \cos{\lambda t} .
\label{eq:ae-formula}
\eeq

Then, putting~\eqref{eq:a-even} into~\eqref{eq:melnikov} and using complex exponentials, gives
\beq
\Delta E = \frac{\pi\e}{\lambda}  e^{-\lambda} 
\intinf \Bigl(\frac{-4}{(1+t^2)^2} + \frac{2}{1+t^2} \Bigr) e^{i\lambda t}dt.
\eeq
This may be closed in the upper complex plane, where the residues at $t=i$
leads to the final answer:
\begin{equation}
\Delta E = -2 \pi^2 \epsilon e^{-2\lambda}.
\label{eq:Delta_E}
\end{equation}

Note that a Melnikov integral has been evaluated to determine the leading
order change of energy, essentially providing the $O(\e)$ term in an infinite
series expansion of this change.  What was found was actually $O(\e
e^{-\sqrt{\frac{2}{\e}}})$, which is significantly smaller.  Alarmingly, then,
the $O(\e^{2})$ or subsequent terms might dwarf the first term in the
expansion, rendering the Melnikov integral meaningless.  A related phenomenon
was studied by Holmes, Marsden, and Scheurle~\cite{HMS:88}, who studied the
rapidly forced pendulum
$$
\theta''+\sin{\theta}= \e^{p} \sin{\frac{t}{\e}}
$$ 
and were able to show that for $p\ge 8$ the Melnikov integral accurately
measures the exponentially small separatrix splitting.  They were subsequently
able to reduce $p$.  Delshams and Seara then removed this restriction on the
size of the rapid forcing term in~\cite{DS:92}.  We therefore have confidence
that the Melnikov integral correctly measures the energy change.  The
numerical evidence of section~\ref{sec:numerics} is also shown to be in
excellent agreement.

Equation~\eqref{eq:Delta_E} may then be used to find the critical velocity:
\begin{align}
2 \left(\diff{X}{t}\right)^2 &= \abs{\Delta E} = 2 \pi^2 \epsilon e^{-2\lambda} \\
V_{\rm c} \equiv \diff{X}{t} &= \pi \sqrt\epsilon e^{-\lambda}.
\label{eq:Vc}
\end{align}
Recall that $t$ has been scaled by a factor of $\e/2$.  Removing this scaling
gives a critical velocity
\beq 
v_{\rm c} = \frac{\pi \epsilon}{\sqrt 2} e^{-\lambda}.
\label{eq:vc}
\eeq

We may compute output velocity $V_{\rm out}$ for slightly supercritical input
velocity $V_{\rm in}= \pi \sqrt\epsilon e^{-\lambda}(1 + \delta_V)$ using the energy:
$$
2 V_{\rm in}^{2}- \Delta E = 2 V_{\rm out}^{2}
$$
so that
$$V_{\rm out} \sim \sqrt{2\delta_V} V_{\rm c}.$$
This gives the characteristic square root behavior of the curve in 
figure~\ref{fig:v_in_out} to the right of $v_{\rm c}$.

We briefly mention two generalizations of the above Melnikov analysis that
will be exceptionally useful in later sections.  On the first near-heteroclinic orbit, we
assume that no energy resides in $a(t)$.  On subsequent near-heteroclines,
$a(t)$ is actively oscillating, so we first suppose that as $t \to -\infty$,
\beq 
a(t) \sim \frac{2\e\pi e^{-\lambda}}{\lambda} A \cos{\lambda(t-T)}, 
\label{eq:a-limit}
\eeq 
where $A$ and $T$ will be determined later. 
Then, since equation~\eqref{eq:model_b} is linear, the contribution due
to this term merely adds to the contribution already calculated.  As before, only
the even part of $a(t)$ is needed for the calculation.  Thus using
$\cos{\lambda(t-T)}= \cos{\lambda T}\cos{\lambda t} + \sin{\lambda T} \sin{\lambda t}$, the total
change of energy is thus
\begin{equation}
\Delta E = (4A \cos{\lambda T}-2) \pi^2 \epsilon e^{-2\lambda}.
\label{eq:Delta_E_extra}
\end{equation}
Depending on the magnitude and sign of $A\cos{\lambda T}$ the energy
change may be positive or negative

Second, we consider the Melnikov integral computed along the separatrix in the
lower half-plane.  System~\eqref{eq:model} obeys the symmetry
$$ (X,\Dot X, a,\Dot a;t) \to (-X,-\Dot X, -a,-\Dot a;t),$$ so that the
Melnikov integral can be computed directly.  Assuming the limiting
behavior~\eqref{eq:a-limit}, the change of energy is
\begin{equation}
\Delta E = (-4A \cos{\lambda T}-2) \pi^2 \epsilon e^{-2\lambda}.
\label{eq:Delta_E_lower}
\end{equation}

\subsection{The full expansion of $a(t)$}
In later sections, we will need more detailed knowledge of the form of $a(t)$.
By equations~\eqref{eq:a_integral}-\eqref{eq:ae-formula}, 
$$
a = 2 a_{\rm e} - \frac{2\e}{\lambda}\int_t^{\infty} \sin{\lambda(t-\tau)}
\frac{\tau}{1+\tau^2} d\tau.
$$
We obtain the asymptotic expression as $t\to +\infty$ by integrating by parts to obtain 
\beq
a(t)
\sim  \frac{2\e}{\lambda^{2}}\Bigl(\frac{t-t_{1}}{(t-t_{1})^2+1} 
+ O(\lambda^{-2})\Bigr)
    -\frac{2\e\pi e^{-\lambda}} {\lambda} \cos{\lambda (t-t_{1})} .
\label{eq:a_expansion_fwd}
\eeq
Similarly, as $t \to -\infty$,
\beq
a(t)
\sim  \frac{2\e}{\lambda^{2}}\Bigl(\frac{t-t_{1}}{(t-t_{1})^2+1} 
+ O(\lambda^{-2})\Bigr)
\label{eq:a_expansion_bkwd}
\eeq
with no exponentially small oscillatory term.
Here we have re-introduced the dependence of the solution on the symmetry time $t_{1}$
from~\eqref{eq:separatrix}, ignored during the calculation above for transparency of notation.
The algebraically small
terms decay for large $t$, so as $t\to\infty$, it is the exponentially small 
oscillating term that dominates.  However, when we use the method of matched asymptotic expansions, we will assume that $t$ is exponentially large of the appropriate size so that the leading order algebraic term and the oscillation are of the same size.

\section{Solutions near $\abs{X}=\infty$}
\label{sec:near_inf}
In the next two sections we construct matched asymptotic solutions to~\eqref{eq:scaledmodel}
by matching near-separatrix solutions to solutions valid near $\abs{X}=\infty$.
The solution for large $\abs{X}$ may be expanded as a near-saddle approach to the
degenerate saddle points at infinity.  Nearly heteroclinic orbits alternate with near-saddle approaches.   Near-saddle expansions for linear
saddle points are common.  In that case, exponential growth of solutions away from the saddle point matches to  exponential decay of homoclinic orbits.  Finite nonlinear saddle points corresponding to  bifurcations for Hamiltonian systems have been analyzed by Diminnie and Haberman~\cite{DH:02, DH:03} and Haberman~\cite{H:01,H:02}.  In the current work, the nonlinear saddle is at infinity, and we do not believe that such an expansion has been analyzed before. In the present case, solutions in the near-saddle region have finite-time singularities which match to the logarithmic growth of the heteroclinic orbits.  We note from the conservative system~\eqref{eq:scaledmodel} and expansion~\eqref{eq:a_expansion_fwd} that the contribution due to $aF'(X)$ is exponentially small for large $t$, so that to leading order
\beq 
4 \ddot X + U'(X)=0 
\label{eq:leading_ode} 
\eeq 
with the energy given
by~\eqref{eq:E_energy}.  $U(X)$ may be approximated in a neighborhood of
$\pm \infty$ by
$$
U \sim -8 e^{\mp 2X}.
$$ We may then form approximations valid for large $X$ in two different ways
depending on whether the energy $E$ is positive or negative.  Phase portraits of~\eqref{eq:leading_ode}, shown in figure~\ref{fig:phaseplane}, may clarify the
results that follow.

If $E= 2 V^{2}>0$, then the solution of~\eqref{eq:leading_ode} corresponding to the near-saddle approach is given by
\beq
 e^{\pm X}= \pm \frac{2}{V}\sinh{V(t-t_{*})} \text{ as } X\to \pm\infty.
\label{eq:approach_positive} 
\eeq 
The $\pm$ sign on the left side of the equation determines whether $X\to
\pm \infty$, and the sign on the right must be chosen so that $\pm (t-t_{*})$
is positive.  The constant $t_{*}$ is the finite blowup time at which time the
near-saddle approach goes to infinity.  The $V$ in the notation is used
intentionally, as it gives the asymptotic velocity of the near approach to the
saddle.

The solution for the near-saddle approach with $E=-2 M^{2}<0$ is given by
\beq
e^{\pm X}= \frac{2}{M}\cos{M(t-t_{**})}
\label{eq:approach_negative}
\eeq 
which has finite time singularities forward and backward in time when $M(t-t_{**})=\pm \frac{\pi}{2}$ and is symmetric
about the symmetry time $t=t_{**}$.

For large $\abs{X}$, $F(X) \sim \mp 4 e^{\pm X}$, so that from~\eqref{eq:model_b},
$$
\ddot a + \lambda^{2} a \sim \pm 4 \e e^{\pm X}.
$$  
Since $\lambda \gg 1$, the asymptotic expansion of $a(t)$ is given by
\beq
a \sim \pm \frac{4\e}{\lambda^{2}} e^{\pm X(t)} + c_{1} \cos{\lambda(t-t_{1})} + c_{2}\sin{\lambda(t-t_{1})},
\label{eq:a-near-infty}
\eeq
where~\eqref{eq:approach_positive} or~\eqref{eq:approach_negative} may be used depending on the circumstance.  Equation~\eqref{eq:a-near-infty} shows that near the saddle approaches $a(t)$ consists of simple harmonic oscillations about a slowly varying mean (which increases in forward and backwards time toward the finite time singularities), all of which can clearly be seen in the numerical calculations. 
The saddle approach with $E<0$, described in detail in the next section must match backwards in time to~\eqref{eq:a_expansion_fwd}, so that $c_2=0$ and $c_1= - \frac{2 \e \pi e^{-\lambda}}{\lambda}$.  Matching this near-saddle approach for $a(t)$ forward in time shows how this exponentially small oscillation is added as previously stated in~\eqref{eq:a-limit}. 

\section{Construction of solutions near the separatrix}
\label{sec:construction}
We now construct an approximation to the initial value problem for the scaled
model equation~\eqref{eq:scaledmodel} under the assumption that the initial
velocity is subcritical.  To be precise, we consider the ``initial value
problem'' defined by~\eqref{eq:scaledmodel} and~\eqref{eq:init_cond}. We let
$V$ refer to the limiting velocity in the scaled model, and reserve $v$ for
the velocity in the physical variables.  We assume that $V>0$ is less than the
critical value found in~\eqref{eq:Vc}.  Then, we may make the assumption that
$E(t)$ stays exponentially close to 0, its value along the heteroclinic orbit.
$X(t)$ may then be approximated in two different ways, depending on whether
$X$ is near a heteroclinic orbit or $X$ is close to infinity.  These two approximations may
then be connected by their limiting behaviors to give a matched asymptotic
expansion.  When $X$ may be approximated by a
heteroclinic orbit
$$ X \approx \pm \sinh^{-1}{(t-t_{j})},$$ 
where $t_{j}$ is the ``symmetry time'' at which $X=0$ for the $j$th nearly heteroclinic orbit. 
For $\abs{X}$ large,
the solution is given by formulas~\eqref{eq:approach_positive}
and~\eqref{eq:approach_negative}.  The exponentially small part of $a(t)$
contributes to the analysis, as it determines the energy difference between
subsequent approaches to infinity.

\subsection{2-bounce solutions}
The 2-bounce solution can be constructed from the following pieces:
\begin{subequations}
\label{eq:connections}
\begin{enumerate}
\item A near saddle approach to $X=-\infty$ with energy $E_{0}=2 V_{0}^{2}$:
\beq e^{-X}=-\frac{2}{V_{0}}\sinh{V_{0}(t-t_{*})},\label{eq:connect1}\eeq 
with $V_{0} < V_{\rm c}$ as given by~\eqref{eq:Vc}.
\item a heteroclinic orbit with $dX/dt>0$:
\beq \sinh X={t-t_1},  \label{eq:connect2}\eeq 
\item a near saddle approach to $X=+\infty$ with negative energy $E=-2M_{1}^{2}$:
\beq 	e^{X}= \frac{2}{M_{1}}\cos{M_{1}(t-t_{**})}, \label{eq:connect3}\eeq
\item a heteroclinic orbit with $dX/dt <0$:
\beq \sinh X=-(t-t_2),  \label{eq:connect4}\eeq 
\item and a near saddle approach to $X=-\infty$ with positive energy $E=2 V_{2}^{2}$:
\beq e^{-X}=\frac{2}{V_{2}}\sinh{V_{2}(t-t_{***})}.\label{eq:connect5}\eeq 
\end{enumerate}
\end{subequations}
The solution can be summarized as a succession of near-saddle approaches,
connected by heteroclinic orbits.  Since the change of energy between consecutive
near-saddle approaches is given by~\eqref{eq:Delta_E}, we see
\beq
-2M_{1}^{2}-2V_{0}^{2}=-2 \pi^2 \epsilon e^{-2\lambda}.
\label{eq:VandM}
\eeq
We now need to compute the change of energy along the second heteroclinic
connection.  We must first compute the symmetry time $t_{2}$ of the second heteroclinic orbit, which is done via leading order matching of $X(t)$.  The algebraically small components of $a(t)$ can be obtained from $X(t)$ by regular perturbation, and thus match immediately once $X$
satisfies matching conditions.  The separatrix is given by $X=-\sinh^{-1}{(t-t_{j})}$, and
the oscillatory part of $a(t)$ is given by $ -\frac{2\e\pi e^{-\lambda}} {\lambda}
\cos{\lambda (t-t_{1})}$ in backwards time.  Shifting time by $t_{2}$, we arrive at
the energy change computed in~\eqref{eq:Delta_E_lower} with $A=-1$ and
$T=t_2-t_1$. The analytic criterion for a 2-bounce solution is that the energy be
positive after the second heteroclinic transition, i.\ e.
\beq
E_{2} = 2 V_{0}^{2} - 2 \pi^{2}\e e^{-2\lambda} + 
(4 \cos{\lambda(t_{2}-t_{1})} -2)\pi^{2} \e e^{-2\lambda} >0.
\label{eq:Delta_E_2}
\eeq
If $E_{2}<0$, then the energy at this saddle approach is less than zero, and the solution does not escape at this saddle approach.

The large time singularity of the first heteroclinic orbit~\eqref{eq:connect2}:
$$ e^{-X} \sim \half\frac{1}{(t-t_{1})}$$ 
must match the singularity of~\eqref{eq:connect3} as $M_{1}(t-t_{**})\searrow-\pi/2$:
$$ e^{-X} \sim \frac{M_{1}}{2} \frac{1}{M_{1}(t-t_{**})+\frac{\pi}{2}},$$
yielding
$$t_{**}-t_{1}=\frac{\pi}{2M_{1}}.$$
A similar calculation yields
$$t_{2}-t_{**}=\frac{\pi}{2M_{1}}.$$ Combining these gives 
\beq
t_{2}-t_{1}=\frac{\pi}{M_{1}}. \label{eq:timespan} 
\eeq 
Note that this is exactly half the period of a closed orbit with
$E=-2M_{1}^{2}$. Matching~\eqref{eq:connect1} to~\eqref{eq:connect2} yields
$t_{*}=t_{1}$, and matching~\eqref{eq:connect4} to~\eqref{eq:connect5} yields
$t_{***}=t_{2}$.  

\subsection{The 2-bounce resonance and the width of the 2-bounce window}
This does not suffice to determine resonant values of
$V_{0}$, because we still need to satisfy the condition that the oscillatory
component of $a(t)$ vanishes in component 5 of the solution.  Thus, at this stage we require a matching condition on the exponentially small oscillating part of $a(t)$.
Two-bounce resonant
solutions are defined by the condition that $E_{2}=2 V_{0}^{2}$.  From~\eqref{eq:Delta_E_2}, 
this requires that $\cos{ \lambda (t_{2}-t_{1})}=1$.  Using~\eqref{eq:timespan}, we obtain the analytic condition for 2-bounce resonant solutions that
$$\frac{\lambda \pi}{M_1}=2\pi n,$$ 
where
$n>0$ is an integer, so that $\Delta E = 2 \pi^2 \epsilon e^{-2\lambda}$.  Thus,
the second jump in energy exactly cancels the first, and all of the energy is
returned to the propagating mode $X$.  This gives a quantization condition
\beq
M_1= \frac{\lambda}{2 n}. \label{eq:quantization}
\eeq 
We can combine this with equation~\eqref{eq:VandM}, to obtain a formula for the initial velocity
of the 2-$n$ resonant solution
\beq
V_{n}=\sqrt{\pi^2 \epsilon e^{-2\lambda}-\frac{\lambda^{2}}{4n^{2}}}.
\label{eq:Vn}
\eeq
$V_n$ denotes the (scaled) initial velocity of the soliton in 2-$n$ resonance with the defect mode.
In order that for $V_{n}$ to be well-defined, $n$ must satisfy 
\beq 
n \ge n_{\rm min}(\e) \equiv \frac{\lambda e^{\lambda}}{2\pi\sqrt{\e}}.
\label{eq:min_bounces}
\eeq
This gives a lower bound on the number of $a$-oscillations in a 2 bounce
resonance, and explains why the observed resonance windows disappear as $\e$ is
increased. 

We may find the width of the 2-$n$ resonance window as follows.  If the energy
change along the second heteroclinic orbit satisfies $\Delta E > 2 M_1^{2}$,
then the solution has positive energy, the trajectory crosses the separatrix,
and the soliton escapes.  If $\Delta E < 2 M_1^{2}$, then the solution remains
bounded, and will approach minus infinity before turning around another time.  Therefore,
the boundaries of the 2-$n$ window, as a function of $M_1$ are given by the
values of $M_1$ where 
$$
\Delta E = 2 M_1^{2}
$$
in~\eqref{eq:Delta_E_2}, i.e.\ if
$$ 
\cos{\frac{\lambda\pi}{M_1}}= \half(1+\frac{M_1^{2}}{\pi^{2}\e e^{-2\lambda}}).
$$
Letting $M_1=\frac{\lambda}{2(n+\delta)}$, then
\beq 
\cos{2 \pi (n+\delta)}= \cos{2 \pi \delta}= \half
\left(1+ \frac{n_{\rm min}^{2}(\e)}{(n+\delta)^{2}}\right).
\label{eq:cos2pid}
\eeq
Considering first the width of the leftmost window, we let $n=\intg{(n_{\rm min}(\epsilon))}+1$, then 
$\delta^{2}= \frac{1}{2n\pi^{2}}(1-\fr{(n_{\rm min}(\epsilon))})$, where $\intg{(Z)}$ and $\fr{(Z)}$
are the integer and fractional parts of $Z$.
Restricting our attention to the smaller windows closer to $v_{\rm c}$, if 
$n\gg n_{\rm min}(\e)$, then $\cos{2\pi \delta} \approx \half$, or
$\delta \approx  \pm \frac{1}{6}.$ The left and right edges of the $n$th
resonance window have velocity approximately
\beq
V_{n\pm}=\sqrt{\pi^2 \epsilon e^{-2\lambda}-\frac{\lambda^{2}}{4(n\pm\frac{1}{6})^{2}}}.
\label{eq:window_edge}
\eeq
If $n$ is sufficiently large, then
$\delta_{n}= \frac{\lambda e^{\lambda}}{2\pi n\sqrt{\e}} \ll 1$, 
and we find that the width of the 2-$n$ window is given by
$$
W_{n}= V_{n+}-V_{n-} \approx V_{\rm c} \delta_{n}^{2} \cdot \frac{3}{n}
$$
which scales as $n^{-3}$ for large $n$.

\subsection{The general initial value problem}
If the second jump in energy, given by by the second Melnikov 
calculation~\eqref{eq:Delta_E_2}, is less than $2 M_{1}^2$, then the soliton 
does not escape on the second interaction with the defect.  Instead it jumps 
to a new energy level inside the separatrix.  We can then replace the 
sequence~\eqref{eq:connections} with the a finite number of nearly heteroclinic orbits separated by near saddle approaches (with negative energy) in which the solution usually escapes at the last saddle approach with positive energy:
\begin{subequations}
\begin{enumerate}
\item A near saddle approach to $X=-\infty$, with energy $E_{0}=2 V_{0}^{2}$:
  \beq
  e^{-X}=-\frac{2}{V_{0}}\sinh{V_{0}(t-t_{*})}\label{eq:genl_connect1}
  \eeq
\item A heteroclinic orbit with $\dot X>0$, over which the change of energy is $\Delta E_{1}$,
  given by the Melnikov integral~\eqref{eq:Delta_E}: 
  \beq 
   \sinh X={t-t_1}  \label{eq:genl_connect2}
  \eeq 
\item A near saddle approach alternating between $X=\pm \infty$, with energy 
   $E_{j}=E_{j-1}+\Delta E_{j}=-2M_{j}^{2}$.
  \beq
   e^{X}= \frac{2}{M_{j}}\cos{M_{j}(t-t_{*}^{j})} \label{eq:genl_connect3}
  \eeq
\item A heteroclinic orbit (alternating between $\dot X<0$ and $\dot X>0$): 
\beq 
\sinh  X=\pm(t-t_j) \label{eq:genl_connect4}
\eeq 
After each nearly heteroclinic orbit, the energy is $E_{j+1}=E_j+ \Delta E_j$.  If $E_{j+1}<0$, the solution
solution has a near saddle approach with negative energy and hence returns to step 3.  However, if $E_{j+1} >0$, the solution escapes, and this last saddle approach is instead mathematically described by step 5.
\item If the solution escapes (at velocity $\pm V_f$), then the near saddle approach at $x=\pm  \infty$ satisfies:
\beq 
e^{\pm  X}=\frac{2}{V_{\rm f}}\sinh{V_{\rm f}(t-t_{***})}\label{eq:genl_connect5}
\eeq
\end{enumerate}
\end{subequations}

Usually the solution will escape after a finite number of bounces.  However, for a set of initial velocities of zero measure, the solution will consist of an infinite number of nearly heteroclinic orbits, will not escape, and will be chaotic. The interesting dynamics take place at step 3 above.  We must again consider the oscillatory part of $a(t)$. In analogy with expansion~\eqref{eq:a_expansion_fwd}, after $j$ near-heteroclinic orbits,
$a(t)$ may be written
\beq
a(t) \sim \text{ algebraically small terms } + 
\frac{2\e\pi e^{-\lambda}} {\lambda}\sum_{k=1}^{j} (-1)^{k+1}\cos{\lambda (t-t_{k})} 
\eeq
where we find $t_{k}-t_{k-1}=\frac{pi}{M_{k-1}}$, the appropriate generalization of~\eqref{eq:timespan}.
The change in energy along the $k$th heteroclinic orbit is given by a generalization of equations~\eqref{eq:Delta_E_extra}--\eqref{eq:Delta_E_lower} to include multiple oscillating terms.
If the solution contains exactly $m$ heteroclinic connections, then the change of energy over
all of the connections is given by the sum of the contributions over all the $m$ nearly heteroclinic orbits, which, after some algebraic manipulation, is
\beq
\Delta E_{\rm total} = \frac{2\e\pi e^{-\lambda}}{\lambda}\sum_{i=1}^{m}\sum_{j=1}^{m}(-1)^{i+j+1}\cos{\lambda(t_{j}-t_{i})}.
\eeq
The condition for an $m$-bounce resonance is thus that $\Delta E =0$, which will happen only for a measure zero set of initial velocities $V_{0}$.  If this is the case,
then $X(t)$ will have interacted with the defect a total of $m$ times.
Between each pair of bounces, $a(t)$ will have undergone an integer number of
complete oscillations (plus a small phase shift).  We may thus construct, in a
manner similar to that above, the $m$-$(q_{1},q_{2},\ldots, q_{m-1})$ bounce
window.  Of course  many of windows do not contain a complete resonance, i.e.\ there does not exist a velocity in the window for which all energy is returned to the propagating mode.  When all the windows 
of initial conditions that eventually escape to $\pm \infty$ are removed, what remains is a Cantor-like set of initial conditions that are trapped for all positive time.

\subsection{The 3-bounce resonance}
\label{sec:3bounce}
\begin{figure}
\begin{center}
\includegraphics[width=4in]{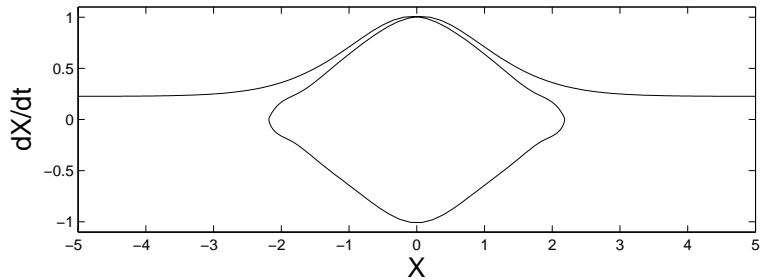}
\caption{A phase-plane portrait of a 3-bounce resonant solution of
  equation~\eqref{eq:scaledmodel}.}
\label{fig:3bounce_phaseplane}
\end{center}
\end{figure}

\begin{figure}
\begin{center}
\includegraphics[width=2in]{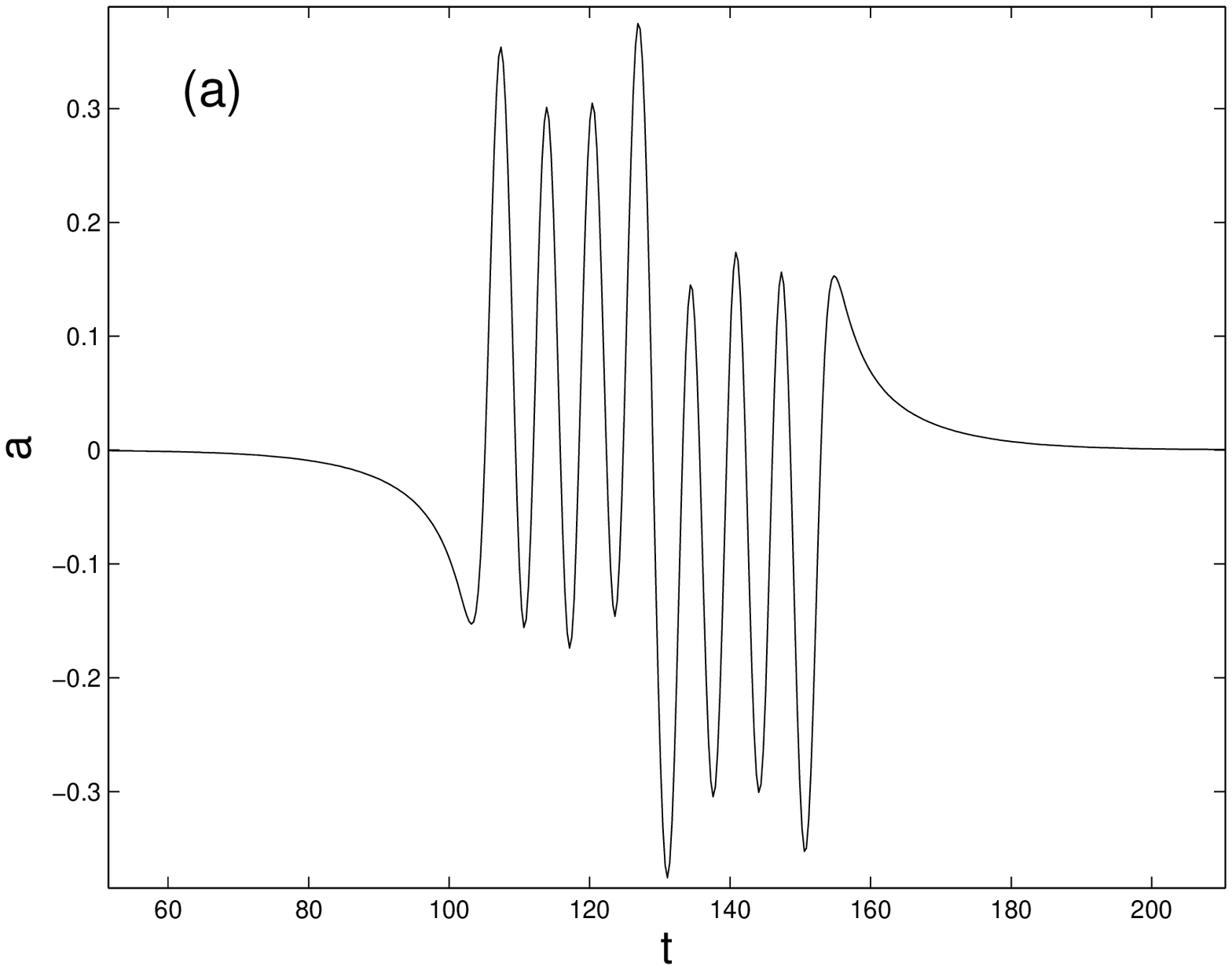}
\includegraphics[width=2in]{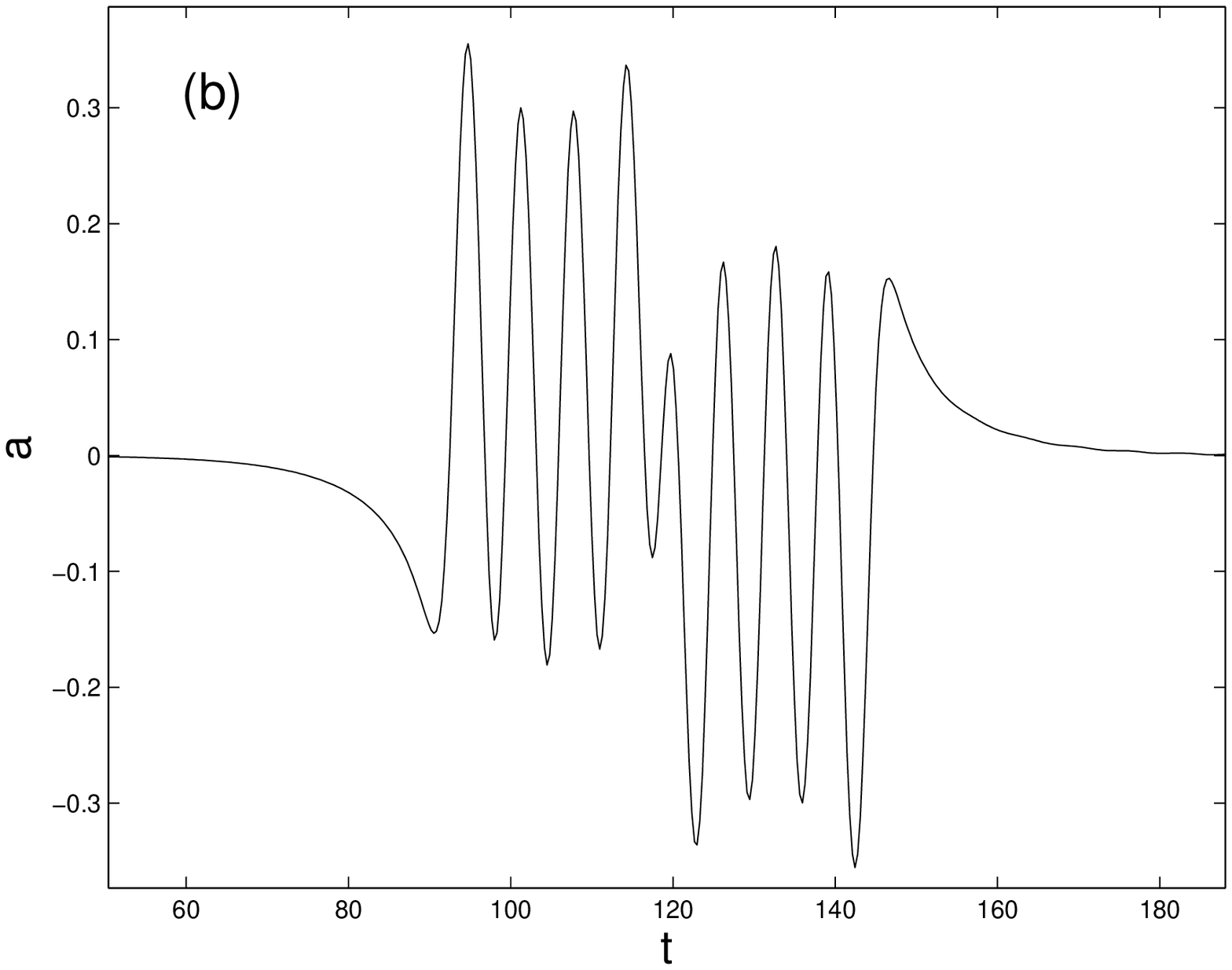}
\caption{The two 3-bounce resonant solutions ($a(t)$ only) to the left and
  right of the first 2-bounce window in figure~\ref{fig:v_in_out}.  In (a),
  $v_=0.09796$.  In (b), $v=0.11301$.}
\label{fig:3bounce}
\end{center}
\end{figure}

It is also possible to construct the three-bounce resonance solutions, which
look in phase space like figure~\ref{fig:3bounce_phaseplane}.
Note that our matched asymptotic expansion depends on $\abs{X}\gg 1$, but this
figure shows that $X \approx 2$ is sufficient.  Although such resonance
windows are too narrow to see with the naked eye in figure~\ref{fig:v_in_out},
careful examination of the data, and use of symmetries allows us to discover
the three-bounce resonance windows.  Note that the two-bounce solutions
consist of $X$ and $a$ which are even functions of $t$ (with the time origin
shifted to be the midpoint between the two singularity times).
Similarly~\eqref{eq:scaledmodel} admits solutions in which both $X(t)$ and
$a(t)$ are odd.  A three bounce resonant solution is an odd function of time,
in which there are three energy jumps and $a(t) \to 0$ as $\abs{t}\to \infty$.
We may assume that the three singularity times are $-t_{0}$, $0$, and $t_{0}$.
Then we note that for the solution to be odd, the energy level $E_{1}$ for $t
\in (-t_{0}, 0)$ must be the same as the energy level $E_{2}$ for $t \in
(0,t_{0})$, so $\Delta E =0$ along the second heteroclinic orbit, i.e.  \beq
\Delta E = (4 \cos{\lambda t_{0}} -2 ) \pi^{2}\e e^{-2\lambda} =0.  \eeq
Therefore $\cos{\lambda t_{0}}=\frac{1}{2}$ or
$$
\lambda t_{0}= 2n\pi \pm \frac{\pi}{3}.
$$
By our standard reasoning this gives
$$
V=\sqrt{\pi^2 \epsilon e^{-2\lambda}-\frac{\lambda^{2}}{4(n\pm\frac{1}{6})^{2}}}
$$ 
which is exactly the formula we obtained in~\eqref{eq:window_edge} when we
ignored a small term in that calculation.  Therefore very close to the edge of
each 2-bounce window, on either side, there exists a symmetric 3-bounce
window. We may check that if before the second energy jump
$$
a(t) \sim -2 \cos{(\lambda t \pm \frac{\pi}{3})},
$$
then afterward
$$
a(t) \sim 2 \cos{(\lambda t \mp \frac{\pi}{3})}
$$ 
so the solution is odd, and we don't need to compute the third interaction.
In figure~\ref{fig:3bounce}, we show the $a(t)$ for the two 3-bounce windows
to the immediate left and right of the first 2-bounce window shown in
figure~\ref{fig:v_in_out}.
Asymmetric 3-bounce windows also exist, in which $a$ oscillates a different
number of times on the first approach to infinity than it does on the second.

\section{Numerical verification}
\label{sec:numerics}
The analysis of the previous section has given us formulas by which we may
compute several features of the solution, as a function of the defect strength
$\e$.  These include, the critical velocity $v_{\rm c}$, the number of
oscillations contained in the solution in the leftmost resonance window
($n_{\rm min}(\e)$), and the locations of the 2-bounce resonance windows, as
well as their widths.
\subsection*{Critical velocities}
\begin{figure}
\begin{center}
\includegraphics[width=3.5inin]{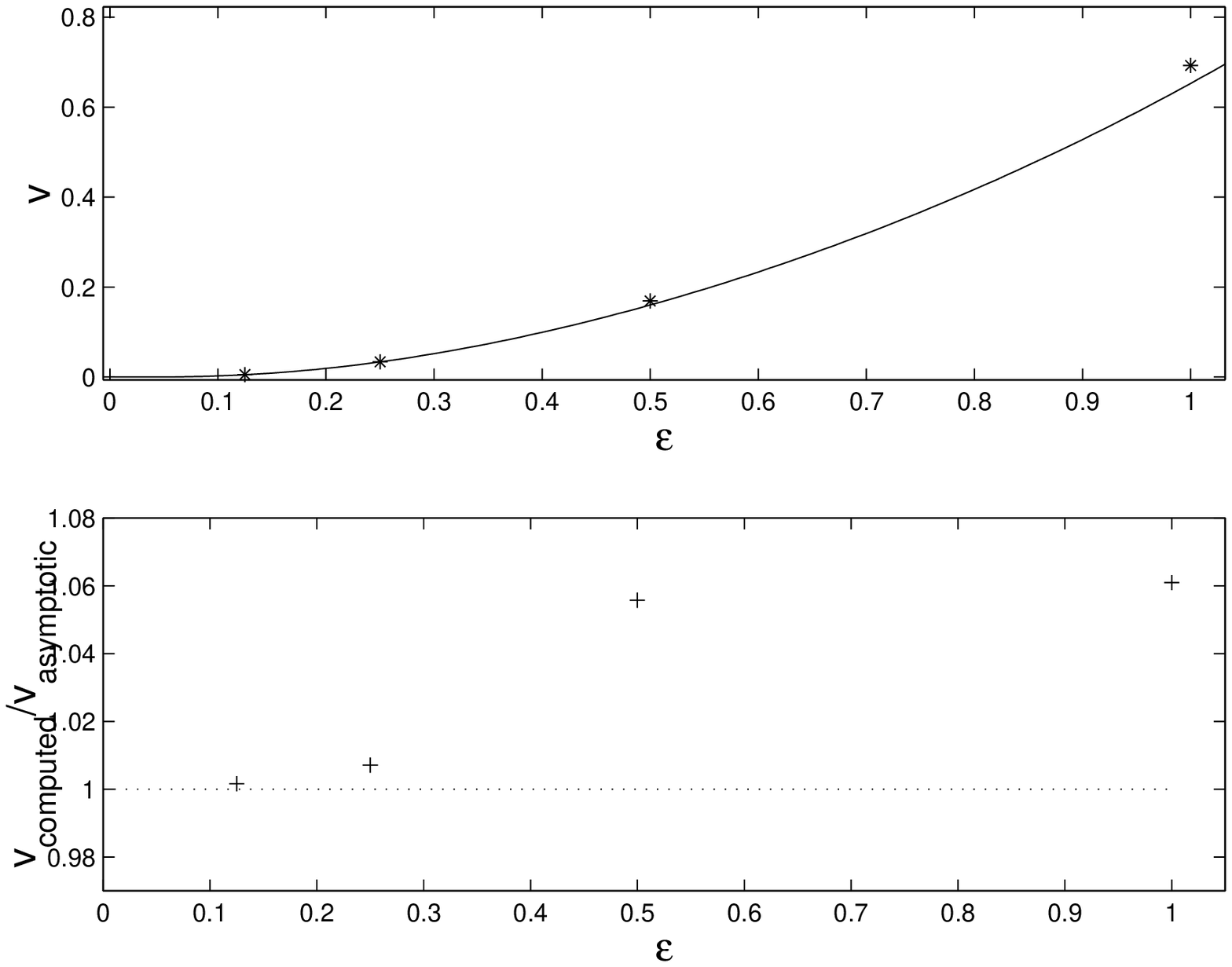}
\end{center}
\caption{(Top) Critical velocity as a function of velocity, numerical $+$, and
via asymptotic calculation (solid line). (Bottom) Ratio of numerical to
asymptotic calculated values.}
\label{fig:epsilon_v}
\end{figure}

Figure~\ref{fig:epsilon_v}, shows the numerically computed critical velocities
for the values $\epsilon \in \{\frac{1}{8},\frac{1}{4},\frac{1}{2},1\},$ as
well as $v_{\rm c}=\pi \e \exp{-\lambda}/\sqrt{2}$.  Of course, both the curve of
calculated velocities, as well as the numerically computed velocities approach
zero as $\e\to 0$, so we must show they approach zero at the same rate to
validate our theory. The lower half of the figure shows the ratio of the
numerical and asymptotic values, which are correct to within 6\% for $\e=1$
and to within 0.2\% for $\e=1/8$.

\subsection*{Predicted minimum $a$-oscillations for resonance ($n_{\rm min}(\e)$)}

For the values $\e = \{\frac{1}{4},\frac{1}{2},1\}$,
formula~\eqref{eq:min_bounces} yields $n_{\rm min}(\e)$ (rounded up to the
nearest whole number: $n_{\rm min}(\frac{1}{4})=15$, $n_{\rm
  min}(\frac{1}{2})=4$, and $n_{\rm min}(1)=1$, which are precisely the values
found via numerical experiment.  The formula gives $n_{\rm
  min}(\frac{1}{8})=98$.  The fewest oscillations seen in the numerical
experiments with $\e=\frac{1}{8}$ was 100, but the equations are very stiff
when $\e$ and $\vin$ are very small, and smaller values of $\vin$ were not
investigated.

\subsection*{Resonance windows}

\begin{figure}
\begin{center}
\includegraphics[width=3.5in]{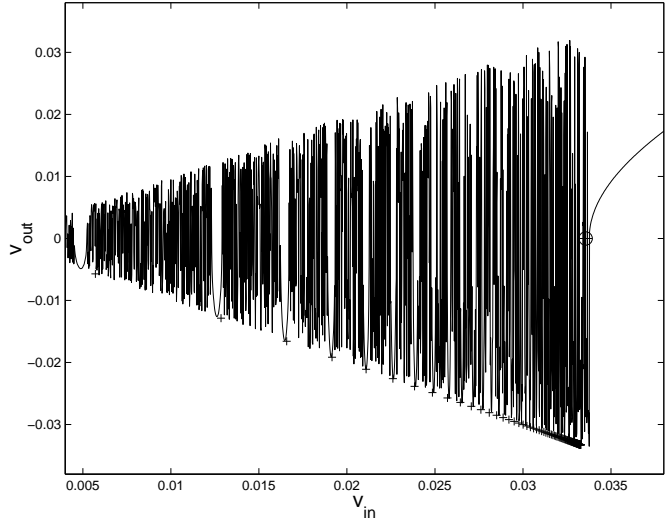}
\end{center}
\caption{Input vs. output velocities for $\e=1/4$ showing the predicted resonant initial velocities
$+$ and the critical velocity $\circ$.}
\label{fig:v_in_out_quarter}
\end{figure}

\begin{figure}
\begin{center}
\includegraphics[width=3.5in]{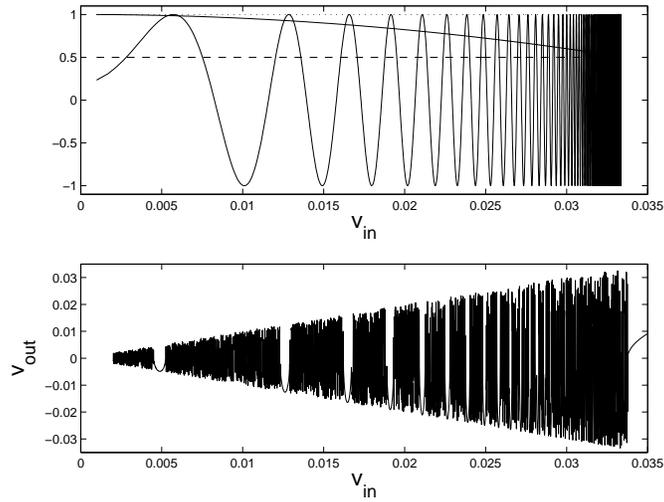}
\end{center}
\caption{(Upper) The oscillatory curve is $y=\cos{2 \pi n(\vin)}$ as a function of
$\vin$.  Its intersections with the line $y=1$ (dotted) give the location of the
2-bounce resonant initial velocities. Intersections with the curve $y = \frac{1}{2} \left(1+ \frac{n_{\rm min} ^{2}(\e)}{n(v)^{2}}\right)$ (solid) give the edges of 
the resonance windows. Intersections
with $y=1/2$ (dashed) give the 3-bounce resonant window velocities.  (Lower) The curve
$\vout$ vs.\ $\vin$.  }
\label{fig:cosnv}
\end{figure}

The comparison of $v_N$ with numerically computed values is shown in
Figures~\ref{fig:v_in_out_quarter} for $\e=1/4$.  Many of the resonance
windows are well-predicted.  We may gain more insight by considering
equation~\eqref{eq:Vn} as defining $n$ as a function of $V$ (and hence as a
function of the unscaled velocity $v$).  In figure~\ref{fig:cosnv} we plot
$\cos{2\pi n(v)}$ as a function of $v$.  If $n \in \mathbb{Z}$, then
$\cos{2\pi n}=1$. Therefore the 2-bounce resonance window centers (i.\ e.\ the resonant 
initial velocities) are given by
the points where the curve $y= \cos{2 \pi n(v)}$ is tangent to the line $y=1$. 
Equation~\eqref{eq:cos2pid} (with $n+\delta$ replaced by $n(v)$) gives the edges 
of the resonance windows.  Therefore to the immediate left and right of the resonance window centers, the curve $y=\cos{2\pi n(v)}$ crosses the curve $y = \frac{1}{2} \left(1+ \frac{n_{\rm min} ^{2}(\e)}{n(v)^{2}}\right)$, giving the window edges. We
note from the figure that this implies that the leftmost resonance windows
should be narrowed with respect to the space between windows.  This is
confirmed in the plot of $\vout$ vs.\ $\vin$.    
Finally, the reasoning of
section~\ref{sec:3bounce} shows that the center of the 3-bounce windows should
be given by the intersection of the curve $y=\cos{2\pi n(v)}$ with the line $y=1/2$.

\section{Conclusions}
\label{sec:summary}
We have shown how a resonant exchange of energy between a soliton and defect
mode gives rise to two bounce resonance windows.  This was known to Campbell
et al.\ as well as to Fei et al.  However by applying perturbation techniques
to a variational model of the system, we have been able to quantify this
effect without recourse to statistical data fitting.  The study of Fei,
Kivshar, and V\'azquez shows remarkable fits between the numerically
determined locations of the resonance windows, and also gives an implicit
equation for the critical velocity that is asymptotically equivalent to our
equation~\eqref{eq:vc}.  The chief advantage of our method is that we are able
to determine the dependence of all these quantities on $\e$ explicitly.

One of us has previously studied the model~\eqref{eq:model} in~\cite{GHW:02}.
In that paper, an artificial coupling parameter $\mu$ is added
to~\eqref{eq:model}.  For small values of $\mu$, we were able to show the
dynamics contained a Smale horseshoe.  In that construction, capture was
identified with transfer of phase space between the regions of
figure~\ref{fig:phaseplane} via turnstile lobes in a certain Poincar\'e map.
That Poincar\'e map was ill-defined as $\mu \to 1$, so the results were not
directly applicable to equation~\eqref{eq:model}, although were very
suggestive of the dynamics.  It does indicate how the dynamics in the regions
between the resonance windows in figure~\ref{fig:v_in_out} depends sensitively
on the input velocity.  Combining this with the quantitative information
contained in the current study gives a rather complete picture of the
dynamics.

Other studies of the 2-bounce resonant phenomenon have often derived a formula for the resonance windows of the form 
$$
\left(v_{\rm c}-v_{n} \right)^{-\half} \sim n T + \theta_{0}
$$ 
where $T$ is the period of the fast oscillations, and $\theta_{0}$ is some
offset time.  The equivalent statement in this study is given in
equation~\eqref{eq:quantization}.  This is equivalent to setting $\theta_{0}$
to zero.  To attain $\theta_{0}$ we would need to find further terms
in~\eqref{eq:timespan}, the equation for the time between interactions, in
terms of the small energy-derived term $M$.  The leading order term is
$O(M^{-1})$ and symmetries of equation~\eqref{eq:leading_ode} show that the
$O(1)$ term must be zero.  The next term in the series is necessarily
$O(M)$. 

Many similar systems have shown the 2-bounce resonance, and the methods
developed here should be adaptable to such systems.  However the current
system is the simplest to study for several reasons.  First, it depends
explicitly on a small coupling parameter $\e$, and when $\e \to 0$ decouples
into two independent oscillations.  Anninos et al.\ derive a variational model
of the kink-antikink scattering in the $\phi^{4}$ experiments of Campbell et
al.~\cite{AOM:91, CSW:83}.  This model does not depend explicitly on a small
parameter, so an artificial one might need to be introduced.  Since our
formula for $v_{\rm c}$ is correct to within 6\% even with $\e=1$, this may be a
reasonable step to take.  Other models do not decouple so cleanly
as~\eqref{eq:scaledmodel} as $\e \to 0$.  Nonetheless, in many systems it is
possible to draw a diagram similar to figure~\eqref{fig:v_in_out}, so we
believe that a similar mechanism is at work.

\section*{Acknowledgements}
We would like to thank Phil Holmes, Michael Weinstein, Greg Kriegsmann, and
Chris Raymond for helpful discussions.  RG was supported by NSF DMS-0204881
and by an SBR grant from NJIT.

\bibliographystyle{amsplain}
\bibliography{SGperturbation}
\end{document}